\documentclass[draft]{agujournal2019}
\usepackage{url,soul,color,float,lineno,amsmath,amssymb,subfigure}
\usepackage[inline]{trackchanges}
\draftfalse
\journalname{Geophysical Research Letters}

\begin{document}
\title{The First GECAM Observation Results on Terrestrial Gamma-ray Flashes and Terrestrial Electron Beams}

\authors{Y. Zhao\affil{1,2}, J. C. Liu\affil{2,3}, S. L. Xiong\affil{2}, W. C. Xue\affil{2,3}, Q. B. Yi\affil{4,2}, G. P. Lu\affil{5}, W. Xu\affil{6}, F. C. Lyu\affil{7}, J. C. Sun\affil{2}, W. X. Peng\affil{2}, C. Zheng\affil{2,3}, Y. Q. Zhang\affil{2,3}, C. Cai\affil{8}, S. Xiao\affil{9,10}, S. L. Xie\affil{11}, C. W. Wang\affil{2,3}, W. J. Tan\affil{2,3}, Z. H. An\affil{2}, G. Chen\affil{2}, Y. Q. Du\affil{12,2}, Y. Huang\affil{2}, M. Gao\affil{2}, K. Gong\affil{2}, D. Y. Guo\affil{2}, J. J. He\affil{2}, B. Li\affil{2}, G. Li\affil{2}, X. Q. Li\affil{2}, X. B. Li\affil{2}, J. Y. Liao\affil{2}, J. Liang\affil{12,2}, X. H. Liang\affil{2}, Y. Q. Liu\affil{2}, X. Ma\affil{2}, R. Qiao\affil{2}, L. M. Song\affil{2}, X. Y. Song\affil{2}, X. L. Sun\affil{2}, J. Wang\affil{2}, J. Z. Wang\affil{2}, P. Wang\affil{2}, X. Y. Wen\affil{2}, H. Wu\affil{12,2}, Y. B. Xu\affil{2}, S. Yang\affil{2}, B. X. Zhang\affil{2}, D. L. Zhang\affil{2}, F. Zhang\affil{2}, P. Zhang\affil{13,2}, H. M. Zhang\affil{2}, Z. Zhang\affil{2}, X. Y. Zhao\affil{2}, S. J. Zheng\affil{2}, K. K. Zhang\affil{14}, X. B. Han\affil{14}, H. Y. Wu\affil{15}, T. Hu\affil{15}, H. Geng\affil{15}, H. B. Zhang\affil{16}, F. J. Lu\affil{2}, S. N. Zhang\affil{2}, H. Yu\affil{1}}
\affiliation{1}{Department of Astronomy, Beijing Normal University, Beijing 100875, Beijing, China}
\affiliation{2}{Key Laboratory of Particle Astrophysics, Institute of High Energy Physics, Chinese Academy of Sciences, Beijing 100049, Beijing, China}
\affiliation{3}{University of Chinese Academy of Sciences, Beijing 100049, Beijing, China}
\affiliation{4}{School of Physics and Optoelectronics, Xiangtan University, Xiangtan 411105, Hunan, China}
\affiliation{5}{School of Earth and Space Sciences, University of Science and Technology of China, Hefei 230026, Anhui, China}
\affiliation{6}{Electronic Information School, Wuhan University, Wuhan 430072, Hubei, China}
\affiliation{7}{Key Laboratory of Transportation Meteorology of China Meteorological Administration, Nanjing Joint Institute for Atmospheric Sciences, Nanjing 210000, Jiangsu China}
\affiliation{8}{College of Physics and Hebei Key Laboratory of Photophysics Research and Application, Hebei Normal University, Shijiazhuang, Hebei 050024, China}
\affiliation{9}{Guizhou Provincial Key Laboratory of Radio Astronomy and Data Processing, Guizhou Normal University, Guiyang 550001, GuiZhou, China}
\affiliation{10}{School of Physics and Electronic Science, Guizhou Normal University, Guiyang 550001, GuiZhou, China}
\affiliation{11}{Institute of Astrophysics, Central China Normal University, Wuhan 430079, HuBei, China}
\affiliation{12}{School of Computing and Artificial Intelligence, Southwest Jiaotong University, Chengdu 611756, SiChuan, China}
\affiliation{13}{College of Electronic and Information Engineering, Tongji University, Shanghai 201804, Shanghai, China}
\affiliation{14}{Innovation Academy for Microsatellites of Chinese Academy of Sciences, Shanghai 201304, Shanghai, China}
\affiliation{15}{National Space Science Center, Chinese Academy of Sciences, Beijing 100190, Beijing, China}
\affiliation{16}{Key Laboratory of Middle Atmosphere and Global Environment Observation, Institute of Atmospheric Physics, Chinese Academy of Sciences, Beijing 100029, Beijing, China}
\correspondingauthor{S. L. Xiong}{xiongsl@ihep.ac.cn}

\begin{keypoints}
    \item During 9-month observation, GECAM has detected 147 bright TGFs, 2 typical TEBs and 2 special TEB-like events.
    \item With novel detector design, GECAM can effectively classify TGFs and TEBs, and reveal their fine temporal features.
    \item We obtained a very high TGF-lightning association rate ($\sim$80\%) between GECAM and GLD360 in east Asia region.
\end{keypoints}

\begin{abstract}
    Gravitational-wave high-energy Electromagnetic Counterpart All-sky Monitor (GECAM) is a space-borne instrument dedicated to monitoring high-energy transients, including Terrestrial Gamma-ray Flashes (TGFs) and Terrestrial Electron Beams (TEBs). We implemented a TGF/TEB search algorithm for GECAM, with which 147 bright TGFs, 2 typical TEBs and 2 special TEB-like events are identified during an effective observation time of $\sim$9 months. We show that, with gamma-ray and charged particle detectors, GECAM can effectively identify and distinguish TGFs and TEBs, and measure their temporal and spectral properties in detail. A very high TGF-lightning association rate of $\sim$80\% is obtained between GECAM and GLD360 in east Asia region.
\end{abstract}

\section*{Plain Language Summary}
    Terrestrial gamma-ray flashes (TGFs) and Terrestrial Electron Beams (TEBs) represent the most energetic radioactive phenomena in the atmosphere of the Earth. They reflect a natural particle accelerator that can boost electrons up to at least several tens of mega electron volts (MeV) and produce gamma-ray radiation. With novel detection technologies, GECAM is a new powerful instrument to observe TGFs and TEBs, as well as study their properties. For example, it is difficult for most space-borne high-energy instruments to distinguish between TGFs and TEBs. However, we show here that, with the joint observation of gamma-ray and charged particle detectors, GECAM can effectively identify TGFs and TEBs. GECAM can also reveal their fine features in the light curves and spectra.

\section{Introduction}

    Terrestrial Gamma-ray Flashes (TGFs) are submillisecond intense bursts of $\gamma$-rays with energies up to several tens of MeV \cite{TGF_GBM_Briggs2010, TGF_AGL_Marisaldi2010, TGF_AGL_Marisaldi2019}, which was serendipitously discovered by \textit{CGRO}/BATSE in 1991 \cite{TGF_BAT_Fishman1994}. Since then, TGFs have been routinely observed by space-borne instruments, such as BeppoSAX \cite{TGF_SAX_Ursi2017}, RHESSI \cite{TGF_RHE_Grefenstette2009}, AGILE \cite{TGF_AGL_Marisaldi2010}, \textit{Fermi}/GBM \cite{TGF_GBM_Roberts2018} and ASIM \cite{TGF_ASM_Ostgaard2019} during last three decades. TGFs can also be observed by ground-based instruments \cite{TGF_GND_Dwyer2012, TGF_GND_Wada2019, TGF_GND_Belz2020}.

    TGFs observed by these space-borne instruments are widely believed to be produced through the initial upward leader of positive Intracloud (+IC) lightning \cite{TGF_VLF_Lu2010, TGF_VLF_Lu2011}. They are the results of relativistic electrons that produce hard X/$\gamma$-rays through the bremsstrahlung process. These electrons are accelerated in a high electric field by the runaway process \cite{TGF_TEO_Wilson1925} and multiplied by many orders of magnitude through the Relativistic Runaway Electron Avalanche process \cite{TGF_TEO_Gurevich1992, TGF_RSC_Dwyer2005}. Two main models were proposed to explain the production of TGFs. One is the lightning leader model, which involves the acceleration of free electrons under the localized electric field in front of lightning leader tips \cite{TGF_TEO_Moss2006, TGF_LIG_Dwyer2010, TGF_TEO_Celestin2011, TGF_LIG_Celestin2013}. The other one is the Relativistic Feedback Discharge (RFD) model \cite{TGF_TEO_Dwyer2003, TGF_TEO_Dwyer2008, TGF_TEO_Dwyer2012, TGF_TEO_Liu2013}, which considers the feedback processes from positrons and photons in a large-scale electric field region. However, the specific mechanism to produce $\sim10^{17}$ to $10^{19}$ electrons is still an open question \cite{TGF_TEO_Chanrion2010, TGF_RSC_Xu2012, TGF_RSC_Xu2015, TGF_TEO_Skeltved2017}.

    By interacting with atmosphere during propagation, TGF photons can produce secondary electrons and positrons. Then they will move along the Earth's magnetic field line, forming Terrestrial Electron Beams (TEBs) \cite{TEB_BAT_Dwyer2008}, which could be observed by some TGF-detecting instruments \cite{TEB_GBM_Xiong2012, TEB_ASM_Sarria2019, TGF_AGL_Lindanger2020}.

    In this study, the data of Gravitational-wave high-energy Electromagnetic Counterpart All-sky Monitor (GECAM) \cite{GEC_INS_Li2022} are utilized for TGFs and TEBs research. GECAM is a space-based instrument dedicated to the observation of gamma-ray electromagnetic counterparts of Gravitational Waves and Fast Radio Bursts, as well as other high-energy astrophysical and terrestrial transients, such as Gamma-ray Bursts (GRBs) \cite{GRB_REV_Zhang2015}, Soft Gamma-ray Repeaters (SGRs), Solar Flares, TGFs and TEBs.

\clearpage
\section{Instrument and Search Algorithm}

    Since launched in December 2020, GECAM has been operating in low earth orbit with 600~km altitude and 29$^{\circ}$ inclination angle \cite{GEC_INS_Han2020}. GECAM consists of twin micro-satellites (i.e. GECAM-A and GECAM-B) and each of them comprises 25 Gamma-ray Detectors (GRDs) \cite{GEC_INS_An2022} and 8 Charged Particle Detectors (CPDs) \cite{GEC_INS_Xv2021}. Each GRD has a geometric area of $\sim$45~cm$^{\rm{2}}$ (round shape with diameter 7.6~cm) and an on-axis effective area of $\sim$21~cm$^{\rm{2}}$ for 1~MeV gamma-rays \cite{GEC_SIM_Guo2020}, while each CPD has a geometric area of 16~cm$^{\rm{2}}$ (square shape with 4.0~cm$\times$4.0~cm) and an on-axis effective area of $\sim$16~cm$^{\rm{2}}$ for 1~MeV electron \cite{GEC_INS_Xv2021}. Considering different orientations of 25 GRDs and 8 CPDs for each GECAM satellite, total effective area of GRDs and CPDs depend on the incident angle. For the incident direction from GECAM's boresight, total effective area of 25 GRDs is $\sim$440~cm$^{\rm{2}}$ for 1~MeV gamma-rays, while that of 8 CPDs is $\sim$20~cm$^{\rm{2}}$ for 1~MeV electrons. Note that only GECAM-B data are utilized here because GECAM-A has not been able to observe yet \cite{GEC_INS_Li2022}. 

    With LaBr$_{3}$ crystals read out by silicon photomultiplier (SiPM) arrays, GRDs can detect high-energy photons in a broad energy range of $\sim$15~keV to $\sim$5~MeV \cite{GEC_INS_Zhang2022a}. CPDs are designed to detect the charged particles (including electrons and positrons) from $\sim$100~keV to $\sim$5~MeV. The joint observation of GRDs and CPDs can distinguish between gamma-rays and charged particle bursts, e.g. TGFs and TEBs \cite{GEC_SFW_Yun2021}. 

    For GRD, the dead time is $4~\mu$s for normal events and $>69~\mu$s for overflow events (i.e. events with higher energy deposition than the maximum measurable energy). Dead time can lead to fewer observed counts, resulting in an underestimation of TGFs' duration and obscuring short TGFs. Each GRD has two read-out channels: high-gain channel ($\sim$15~keV--$\sim$300~keV) and low-gain channel ($\sim$300~keV--$\sim$5~MeV) \cite{GEC_INS_Liu2021}. The design, performance, and other information about GECAM have been reported by \citeA{GEC_INS_Li2022, GEC_INS_An2022, GEC_INS_Xv2021}.

    The considerable number of GRDs is helpful to locate source region of TGFs. We have proposed a dedicated localization method for all-sky monitor which can be used for extremely short-duration TGFs \cite{YZ_LOC_MTD}. Despite the limited counting statistics of TGFs, GECAM is capable of roughly determining the location of TGF candidates, although the error is large \cite{YZ_LOC_GEC}.

    To detect those extremely short and bright bursts, e.g. TGFs and TEBs, a dedicated anti-saturation data acquisition system (DAQ) is designed for GECAM. The data buffer in DAQ can accommodate up to 4092 and 1020 counts for the high-gain channel and low-gain channel of each GRD, respectively. Since there are usually only several hundreds of counts registered for bright TGFs, GECAM's DAQ can guarantee to transfer and save almost all TGFs counts that are recorded by GECAM detectors \cite{GEC_INS_Liu2021}.

    As the main contamination source for TGFs, cosmic-ray events show very similar patterns in data as TGFs, but with an even shorter duration. Thanks to GECAM's high time resolution, i.e. 100~ns \cite{GEC_CAL_Xiao2022}, GECAM can effectively distinguish between cosmic-ray events and TGFs. Indeed, a dedicated data product called Simultaneous Events is designed for GECAM. The Simultaneous Events Number (SimEvtNum) is defined as events number from different detectors registered in the same 300~ns time window \cite{GEC_CAL_Xiao2022}. As the SimEvtNum increases, the probability of these events caused by cosmic-rays surges. Thus the events marked with SimEvtNum$\geq$13 are not utilized in the searching, as they may be the result of cosmic rays.

    To unveil TGFs and TEBs in GECAM data, we developed a dedicated burst search algorithm, which is different from normal burst search for GRBs \cite{HXM_SEA_Cai2021, HXM_SEA_Cai2023}, because TGFs and TEBs are so weak that only a few counts are registered in each detector, and both GRDs and CPDs are needed in searching. The event-by-event (EVT) data of GECAM GRDs and CPDs are used in this study. Only recommended normal events with SimEvtNum$<$13 are utilized. We divide 25 GRDs into four groups considering the neighboring position, resulting in three groups with six GRDs and one with seven GRDs. All 8 CPDs are treated as a single group.

    Assuming the background follows the Poisson distribution, the probability that the counts are from background fluctuation can be calculated as:

        \begin{align}
            P_{\rm group}(S \geq S^{'}|B) = 1 - \sum_{S=0}^{S=S^{'}-1} \frac{ B^{S} \cdot \exp(-B) }{ S! } ,
            \label{EQ_Search_1}
        \end{align}
    where $S$ and $S^{'}$ are observed counts and threshold counts, respectively, for one group in a time window, $B$ is the estimated background for the time window calculated by the average counts over $T_{\rm rela} \in$ [-5,-1] s and $\in$ [+1,+5] s, where $T_{\rm rela}$ is relative time regarding the end time of the time window.

    For a given search bin, we calculate the joint probability of $N_{\rm{trig}}^{'}$ or more groups out of a total of $M$ groups surpassing the trigger threshold for a single group. This joint probability ($P_{\rm{bin}}$) can be given by:
        \begin{align}
            P_{\rm{bin}}(N_{\rm{trig}} \geq N_{\rm{trig}}^{'}) = \sum_{N_{\rm{trig}} = N_{\rm{trig}}^{'}}^{N_{\rm{trig}} = M} \binom{N_{\rm{trig}}}{M} \cdot ( P_{\rm group} )^{ N_{\rm{trig}} } \cdot ( 1 - P_{\rm group} )^{ M - N_{\rm{trig}} } .
            \label{EQ_Search_2}
        \end{align}

    Here, seven time scales are utilized for searching. The widths of time scales with the corresponding empirical threshold $P_{\rm{bin}}$ are: 50~$\mu$s ($5.0\times10^{-22}$), 100~$\mu$s ($2.0\times10^{-21}$), 250~$\mu$s ($1.3\times10^{-20}$), 500~$\mu$s ($5.0\times10^{-20}$), 1~ms ($2.0\times10^{-19}$), 2~ms ($8.0\times10^{-19}$), 4~ms ($3.2\times10^{-18}$). For instance, we required $\geq$2 GRD groups to have $\geq$8 counts each in a 100~$\mu$s time bin, which corresponds to a P-value of $\sim7.3\times10^{-12}$ for one group with background level of 400 counts/s for one GRD. Considering the joint probability (Equation 2), the P-value for a given search bin was calculated to be $2.0\times10^{-21}$. All time scales are used for TGF search, while only the last four are used for TEB search. These empirical criteria are relatively strict so that only intense TGFs or TEBs could be identified.

    We can derive the trigger threshold for a group of GRDs, $P_{\rm{group,GRD}}$, using $P_{\rm{bin}}$ by setting $M=4$ and $N_{\rm{trig},GRD}^{'}=2$:
        \begin{align}
            P_{\rm{bin}}(N_{\rm{trig}} \geq 2) = 6 \cdot P_{\rm{group,GRD}}^{2} - 8 \cdot P_{\rm{group,GRD}}^{3} + 3 \cdot P_{\rm{group,GRD}}^{4}.
            \label{EQ_Search_3}
        \end{align}

    Similarly, we can derive the trigger threshold for TEBs with CPDs, $P_{\rm{group,CPD}}$, using $P_{\rm{bin}}$ by setting $M=1$ and $N_{\rm{trig},CPD}^{'}=1$:
        \begin{align}
            P_{\rm{bin}} = P_{\rm{group,CPD}} .
            \label{EQ_Search_4}
        \end{align}

    For candidates to be identified as TGFs/TEBs, all criteria below must be met:

    \begin{itemize}

        \item[1.] The trigger threshold (Equations \ref{EQ_Search_3} and \ref{EQ_Search_4}) must be satisfied.

        \item[2.] Candidates should not be SGRs. Note that millisecond-duration SGRs can be searched in the time scale of milliseconds with a much softer spectrum than TGFs.

        \item[3.] Should not be caused by instrument effects, which are characterized by that there is significant excess (Poisson significance $>$6~$\sigma$) registered in 2 to 3 GRDs while no obvious signals (Poisson significance $<$3~$\sigma$) for most (i.e. $>$21) GRDs.

        \item[4.] For filtering out cosmic-rays, ratio of the simultaneous event ($R_{\rm{sim,7}}$\footnote{$R_{\rm{sim,7}}$: total simultaneous events number registered in $>$7 GRDs, divided by total events number in the searching bin.}) should be $<$20\%.

    \end{itemize}

    For the identification of TEBs, more criteria are needed which will be described in Section 4. To further illustrate the capability of GECAM to identify cosmic-rays, a case is illustrated in the supplementary material in Supporting Information.

\clearpage
\section{GECAM TGFs}

\begin{figure*}
        \centering
        \subfigure[]{\includegraphics[width=\columnwidth]{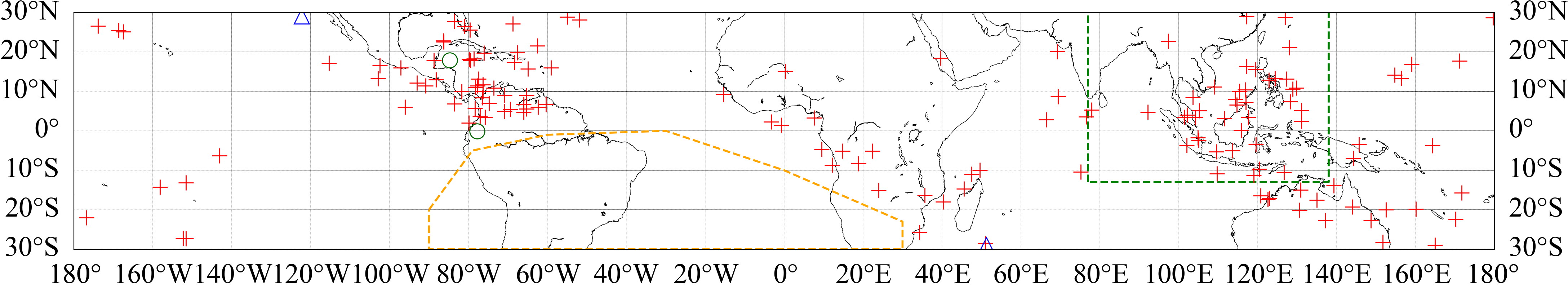}}
        \quad
        \\
        \subfigure[]{\includegraphics[height=6.0cm]{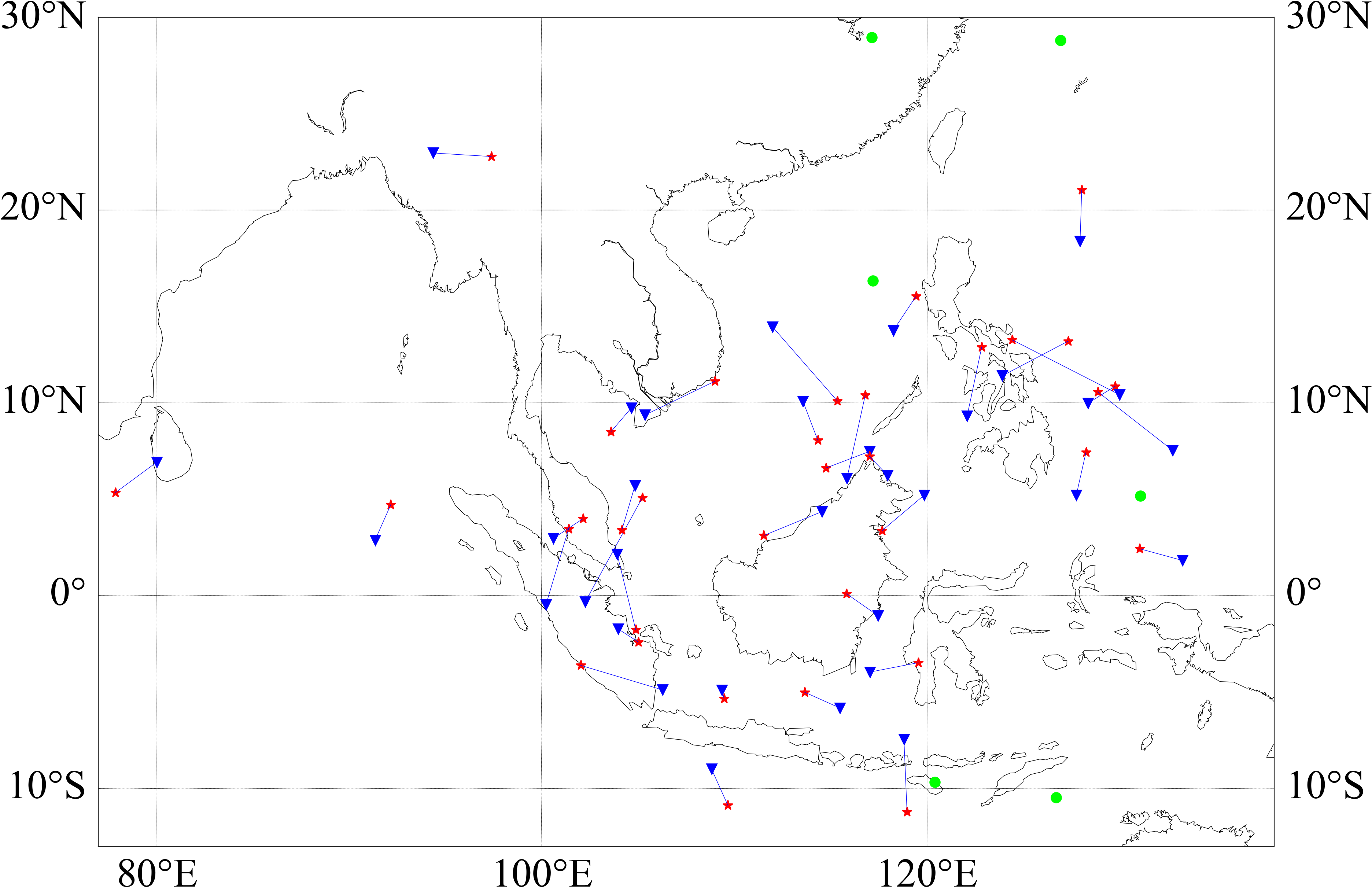}}
        \quad
        \\
        \subfigure[]{\includegraphics[height=4.5cm]{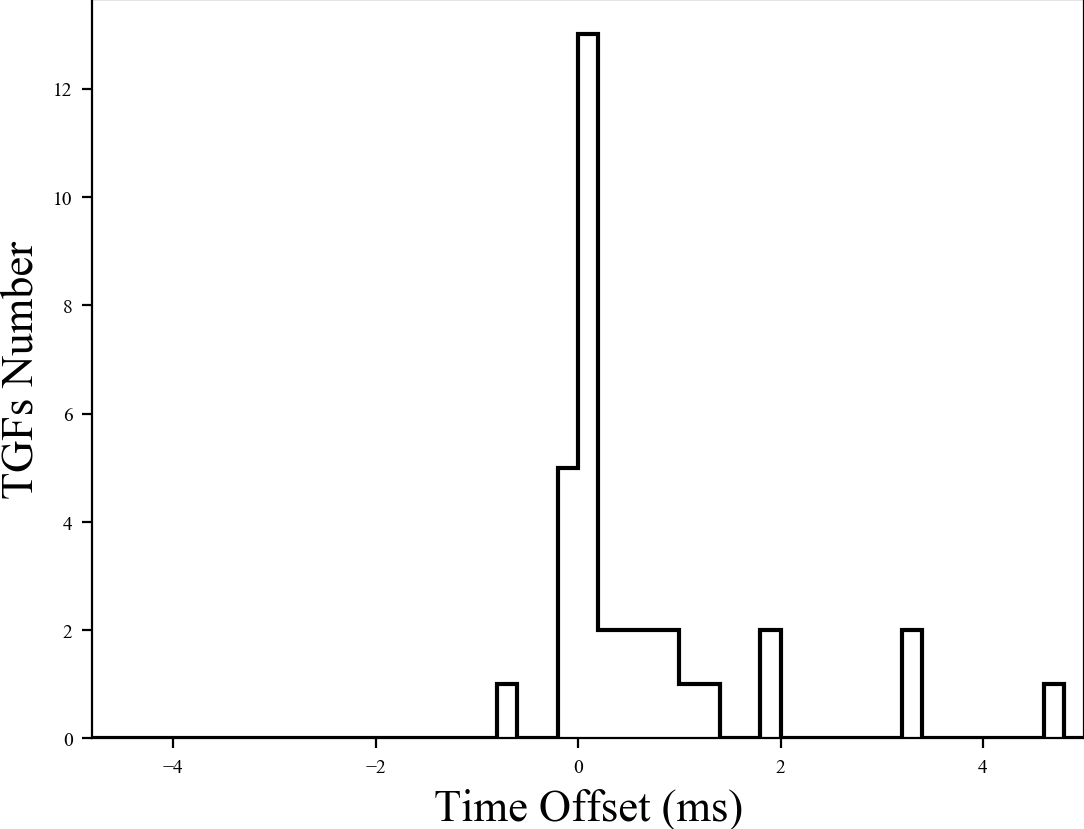}}
        \quad
        \subfigure[]{\includegraphics[height=4.5cm]{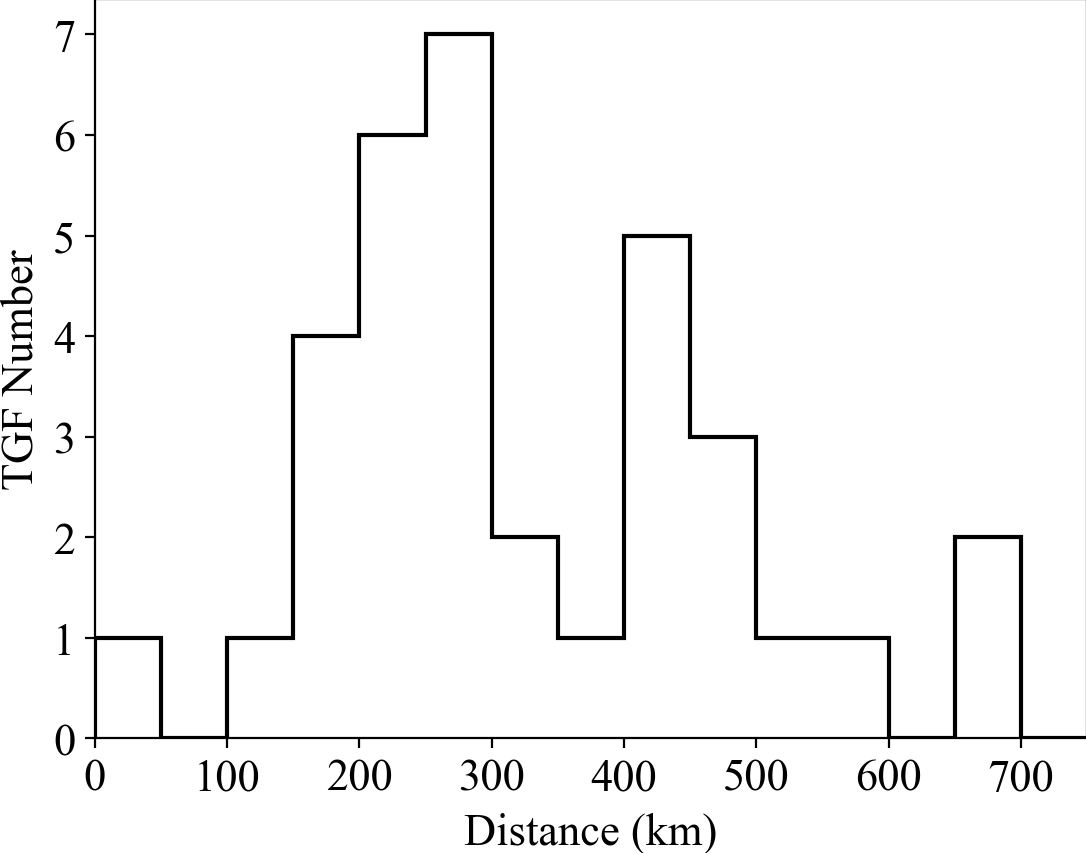}}
        \quad
        \caption{Geographical distribution of GECAM TGFs. (a) GECAM nadirs of 147 TGFs (red pluses), 2 TEBs (green circles) and 2 special TEB-like events (blue triangles, see Section 4). The green and orange dashed lines show the east Asia region (EAR, $77^{\circ}$~E--$138^{\circ}$~E, $13^{\circ}$~S--$30^{\circ}$~N) and South Atlantic Anomaly (SAA), respectively. (b) The red[lime] markers illustrate TGFs with[without] associated GLD360 lightning inside the EAR. The blue triangles illustrate the associated lightning within $\pm$5~ms corrected for light travel time and within 800~km from GECAM nadirs. (c) Distribution of time offsets (corrected for propagation time) between GECAM TGFs and their associated lightning. (d) Distribution of sphere distance between the GECAM nadirs and their associated lightning.}
        \label{FIG_2}
    \end{figure*}

    From December 10th, 2020 to August 31st, 2022, the effective observation time of GECAM-B is $\sim$274.5 days ($\sim$9 months or $\sim$0.75 years). As shown in Figure \ref{FIG_2}, 147 TGFs are identified by our search algorithm, corresponding to a discovery rate of $\sim$200 TGFs/year or $\sim$0.54 TGFs/day. We note that this TGF sample only contains bright ones, resulting from the strict searching threshold. Therefore, GECAM's TGF discovery rate would increase as we decrease the search threshold in the future.

    The Global Lightning Dataset (GLD360) is utilized to match lightning for GECAM TGFs in the time window of $\pm$ 5~ms corrected for light propagation time and within the distance window of 800~km from GECAM nadirs. The GLD360 lightning-association ratio is $\frac{34}{41} \approx 80\%$ in the east Asia region (EAR, $77^{\circ}$~E--$138^{\circ}$~E, $13^{\circ}$~S--$30^{\circ}$~N) which is $\sim$2.5 times of results based on data of the other space-borne instruments and the World Wide Lightning Location Network (WWLLN) lightning ($\sim$33\%) \cite{TGF_GBM_Roberts2018, TGF_AGL_Maiorana2020}. The high lightning-association ratio may be attributed to two factors: (1) The detection efficiency of GLD360 is higher than the other lightning location network \cite{TGF_GLD_Said2013, TGF_GLD_Poelman2013, TGF_GLD_Pohjola2013}. \citeA{TGF_GBM_Mailyan2020} have also confirmed that using GLD360 lightning data significantly improves the association ratio between \textit{Fermi}/GBM TGFs and sferics. (2) This GECAM sample only contains bright TGFs, and their associated lightning strokes maybe brighter. As shown in Figure \ref{FIG_2}c, most of the time offsets (corrected for propagation time) between GECAM TGFs and their associated lightning are centered around $\pm$2~ms. Distances between GECAM nadirs and their associated lightning range from $\sim$50 to $\sim$800~km. These time offsets and distances are consistent with previous reports, although the chance probability is $\sim$2.7\% higher than previous studies using WWLLN dataset \cite{TGF_GBM_Roberts2018} due to the high detection efficiency of GLD360. With 41 TGFs in the east Asia region, there would be $\sim$1.1 false associations. However, if we only consider associations within $\pm$2~ms, the probability of chance associations is $\sim$1.1\%, resulting in only $\sim$0.4 false associations with the 41 TGFs. Since 31 out of 34 lightning events are centered at $\pm$2~ms of TGFs time, we conclude that most of the associated lightning events are genuine matches.

\begin{figure*}
        \centering
        \subfigure[]{\includegraphics[height=4.5cm]{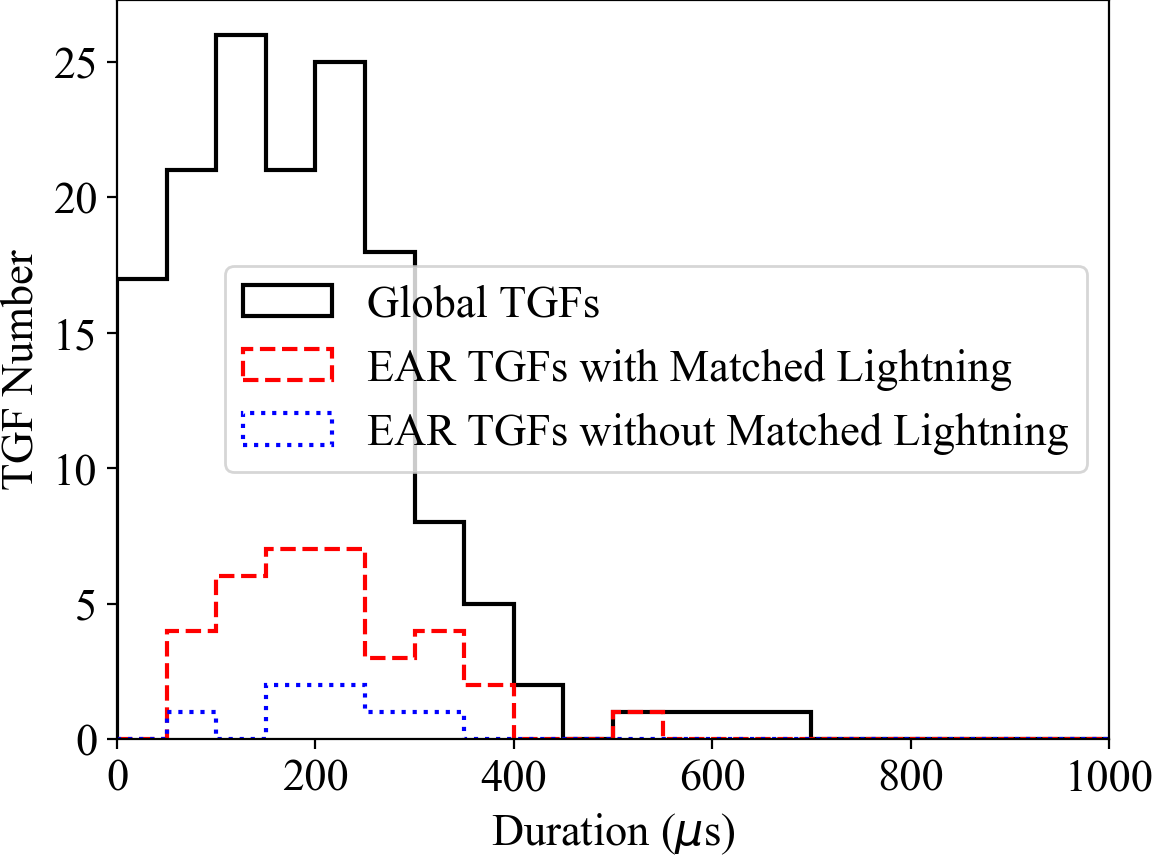}}
        \quad
        \subfigure[]{\includegraphics[height=4.5cm]{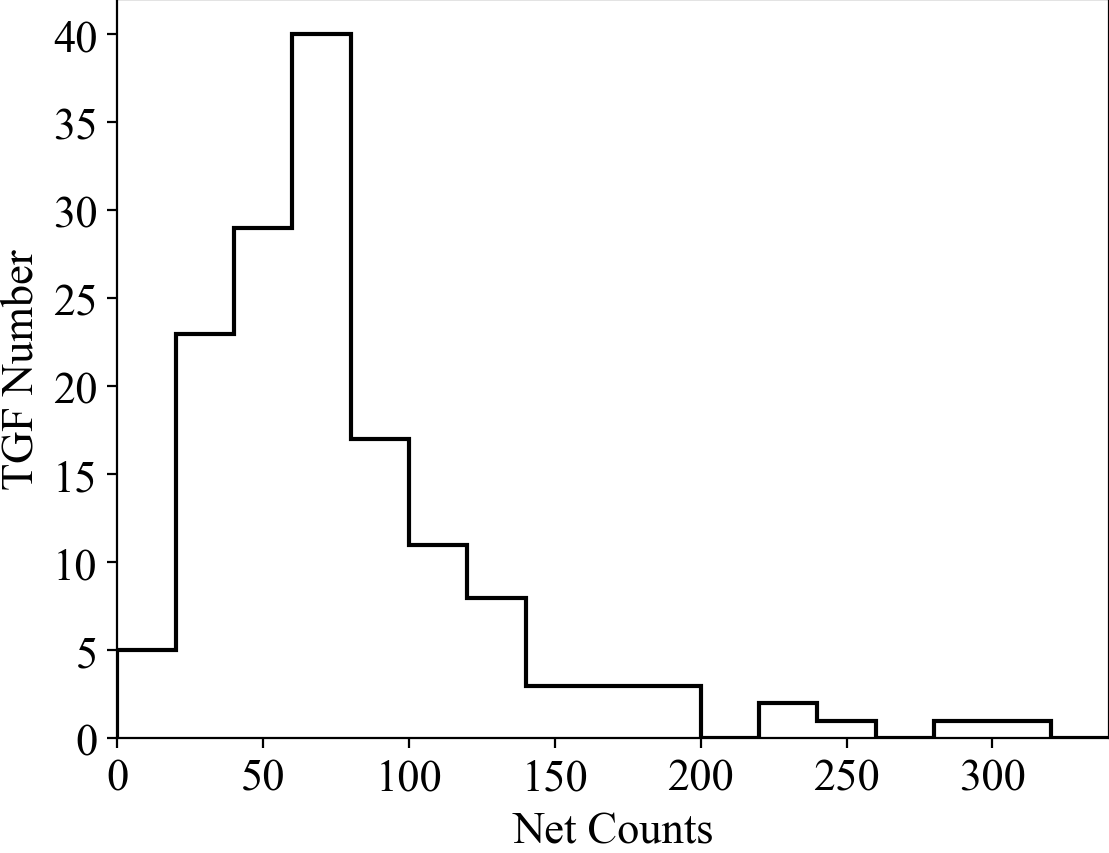}}
        \quad
        \\
        \subfigure[]{\includegraphics[height=5.5cm]{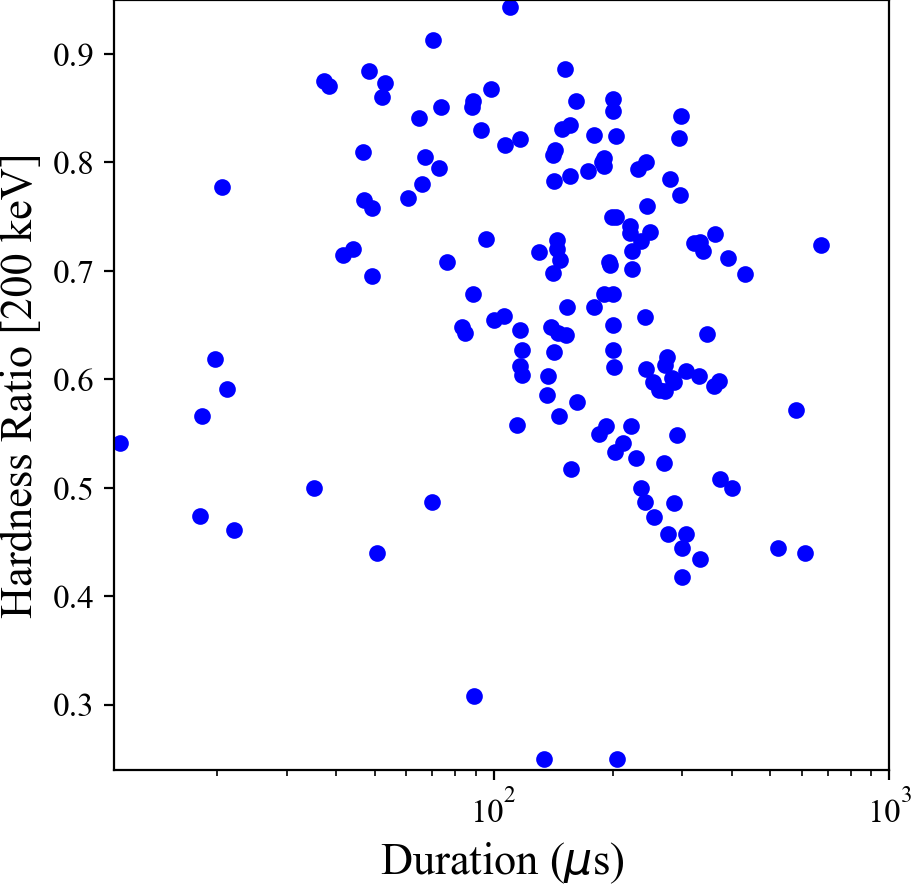}}
        \quad
        \subfigure[]{\includegraphics[height=5.5cm]{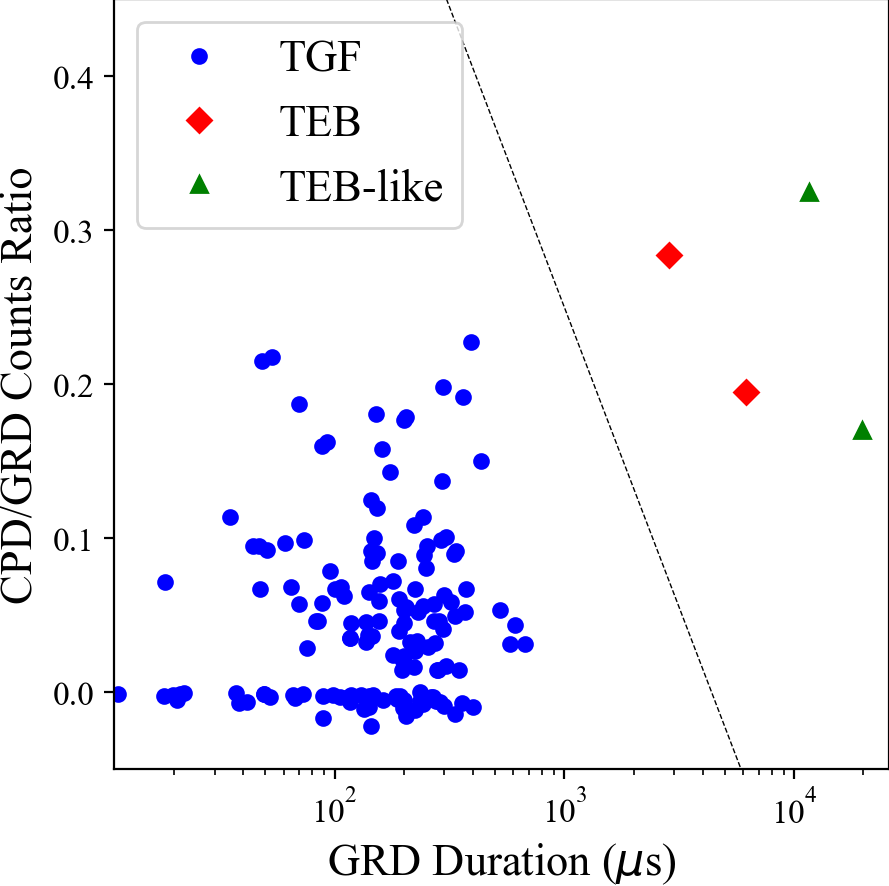}}
        \quad
        \caption{Statistical properties of GECAM TGFs and TEBs. (a) Duration distribution of TGFs. The duration is calculated by the Bayesian Blocks algorithm. The black, red, and blue lines illustrate the duration distribution of total TGFs (147), TGFs with (34), and without (7) associated GLD360 lightning in the EAR, respectively. (b) Distribution of observed net counts for total TGFs. (c) Scatter plot of TGFs' duration versus hardness ratio (energy limitation 200~keV). (d) Scatter plot of duration versus CPD/GRD counts ratio for TGFs (blue circles), TEBs (red diamonds) and TEB-like events (green triangles, see Section 4). The dashed line shows a tentative threshold of equation $y = -0.39\times\log_{10}(x) + 1.42$ for TGF/TEB classification, where x is the duration ($\mu$s) and y is the CPD/GRD counts ratio.}
        \label{FIG_3}
    \end{figure*}

    The statistical distribution of temporal, intensity and energy properties of GECAM TGFs are shown in Figure \ref{FIG_3}. The duration is calculated by the Bayesian Block (BB) algorithm \cite{STAT_BayesianBlock}. The distribution of GECAM TGFs' duration is centered around $\sim$200~$\mu$s (see Figure \ref{FIG_3}a). We note that the proportion of GECAM TGFs with extremely short duration (i.e., $<$40~$\mu$s) is less than that observed by ASIM \cite{TGF_ASM_Ostgaard2019}, which may be due to the strict searching threshold, although different instruments' duration cannot be compared directly. Figure \ref{FIG_3}c demonstrates that TGFs with shorter duration typically exhibit a harder spectrum, which is consistent with previous observations \cite{TGF_GBM_Briggs2013}. The pulse pile-up effect of \textit{Fermi}/GBM can reduce observed counts and make the measured spectrum harder \cite{INS_EFF_Bhat2014}. Similarly, GECAM's pulse pile-up effect also has such an impact. Therefore, this phenomenon might be partly due to the pile-up effect. It appears that there are some TGFs with relatively soft spectra below the diagonal line. However, these TGFs also satisfy that short-duration TGFs have hard spectra. Since the sample number is limited, we will investigate this phenomenon further as the number of sample increases. As shown in Figure \ref{FIG_3}d, the duration and CPD/GRD counts ratio is effective to classify TGFs and TEBs (see Section 4).

\begin{figure*}
        \centering
        \subfigure[$T_{\rm 0}$ UT 2021-02-01T02:09:25.6915]{\includegraphics[height=3.0cm]{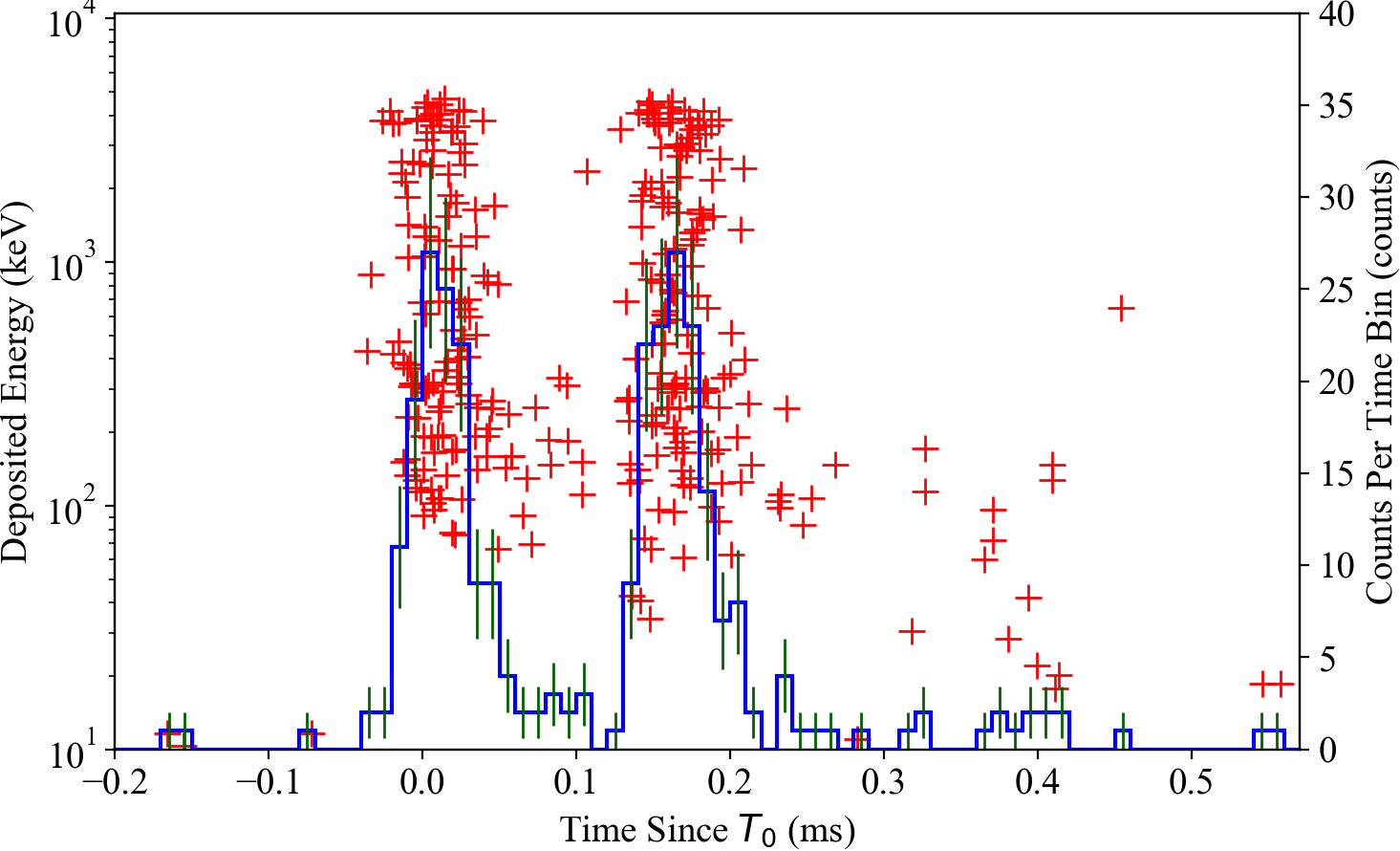}}
        \quad
        \subfigure[$T_{\rm 0}$ UT 2021-07-10T21:19:04.5195]{\includegraphics[height=3.0cm]{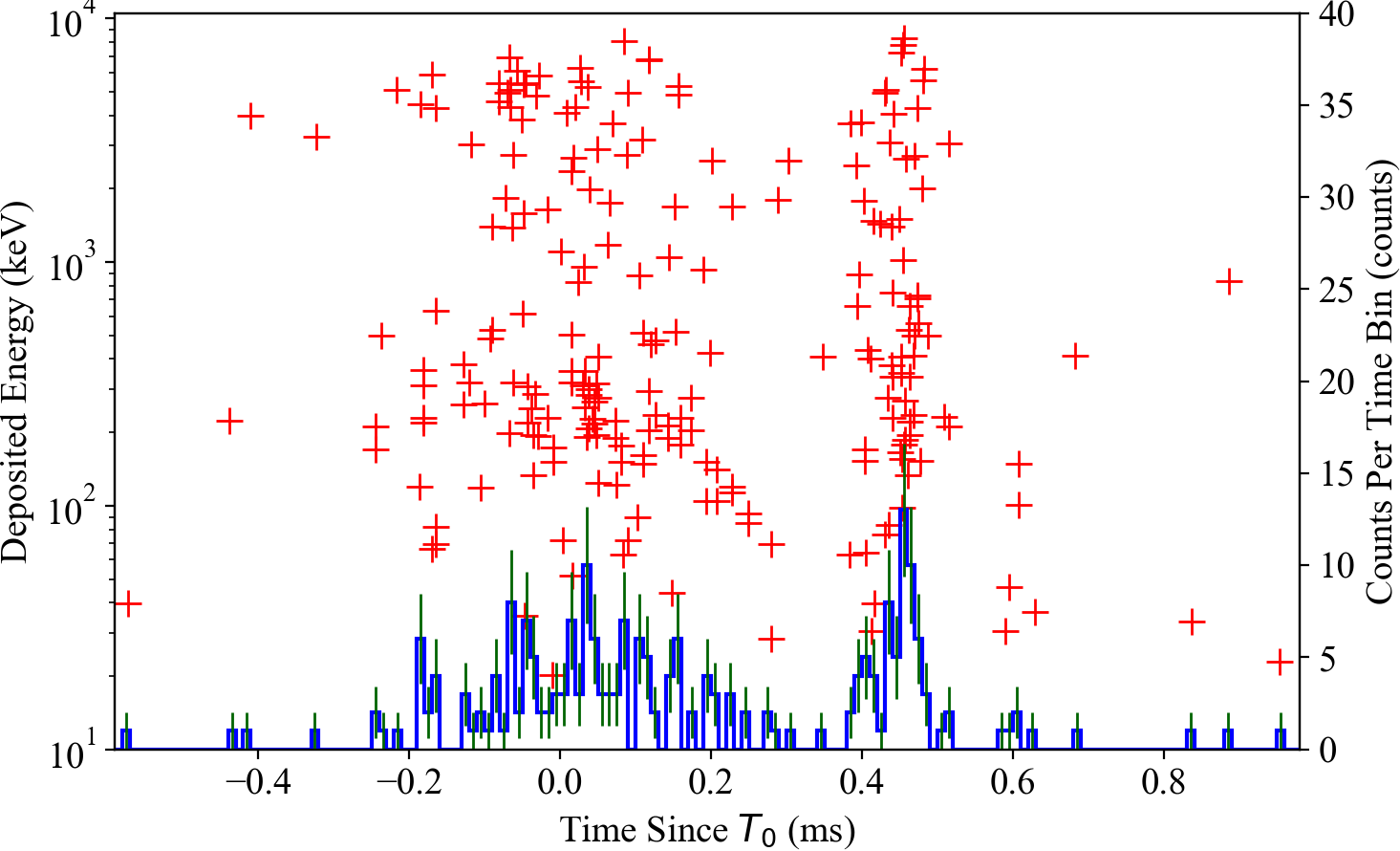}}
        \quad
        \subfigure[$T_{\rm 0}$ UT 2022-01-22T22:24:49.6646]{\includegraphics[height=3.0cm]{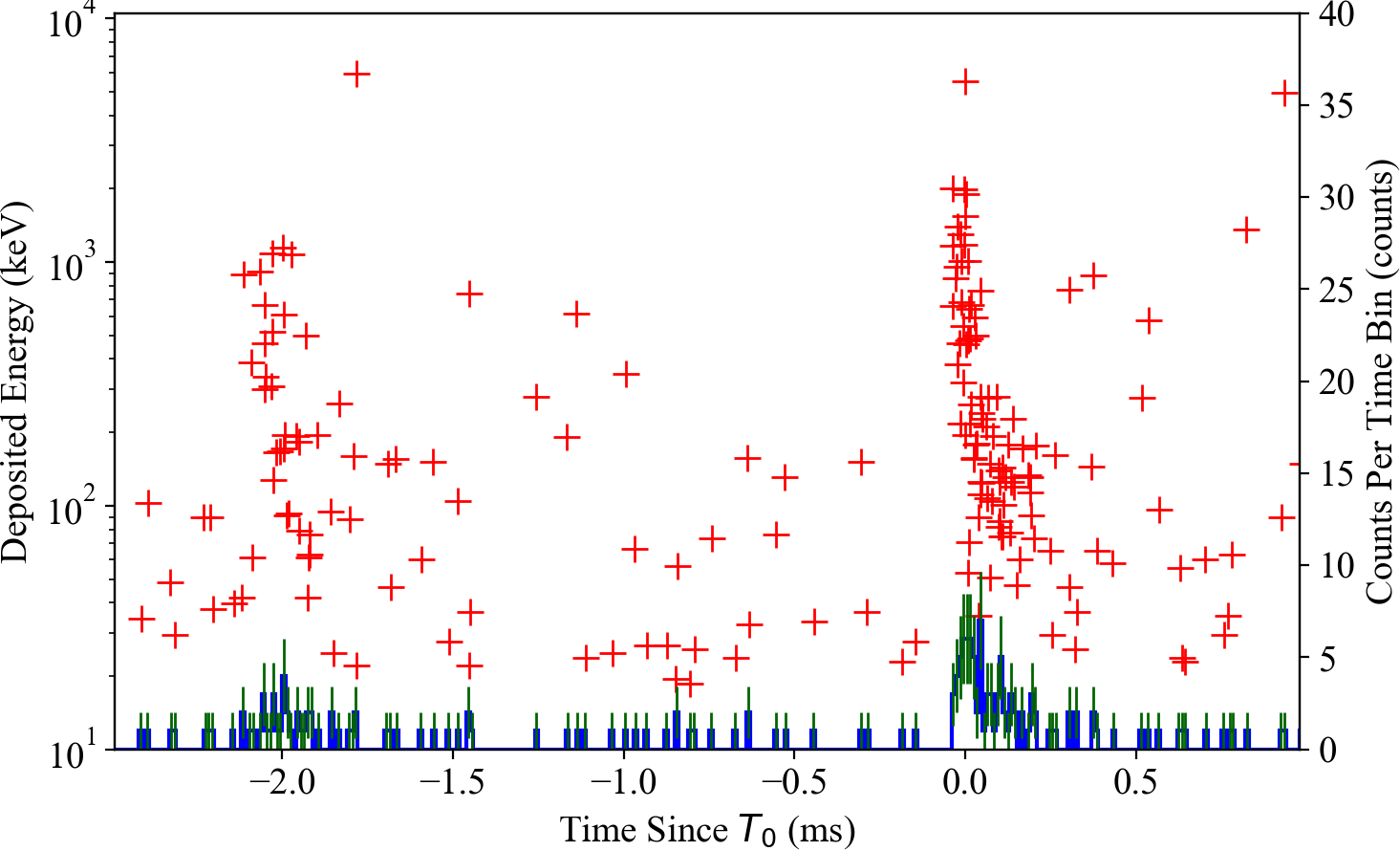}}
        \quad
        \subfigure[$T_{\rm 0}$ UT 2021-03-07T19:13:49.9955]{\includegraphics[height=3.0cm]{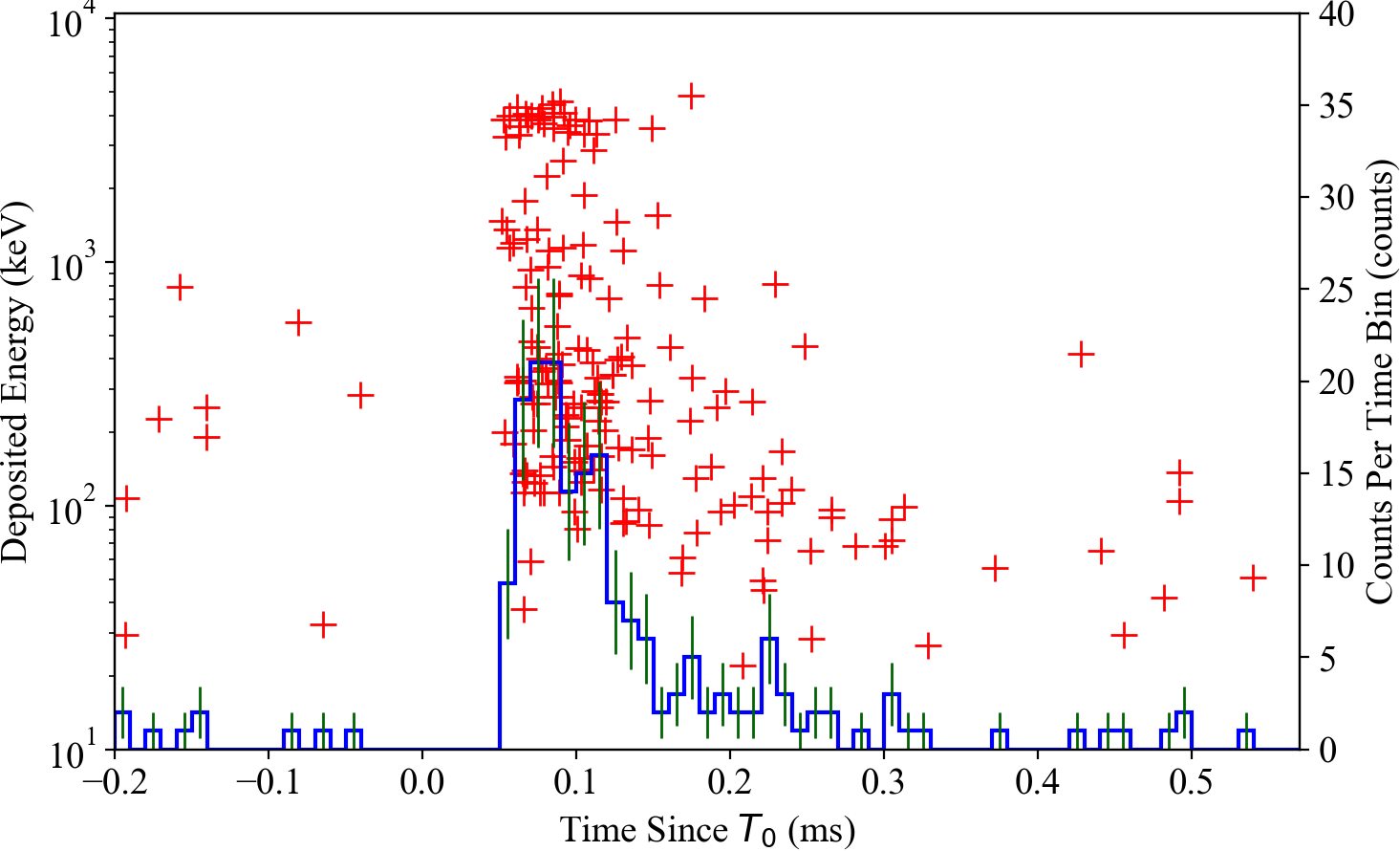}}
        \quad
        \\
        \subfigure[$T_{\rm 0}$ UT 2021-03-29T06:56:37.8318]{\includegraphics[height=3.0cm]{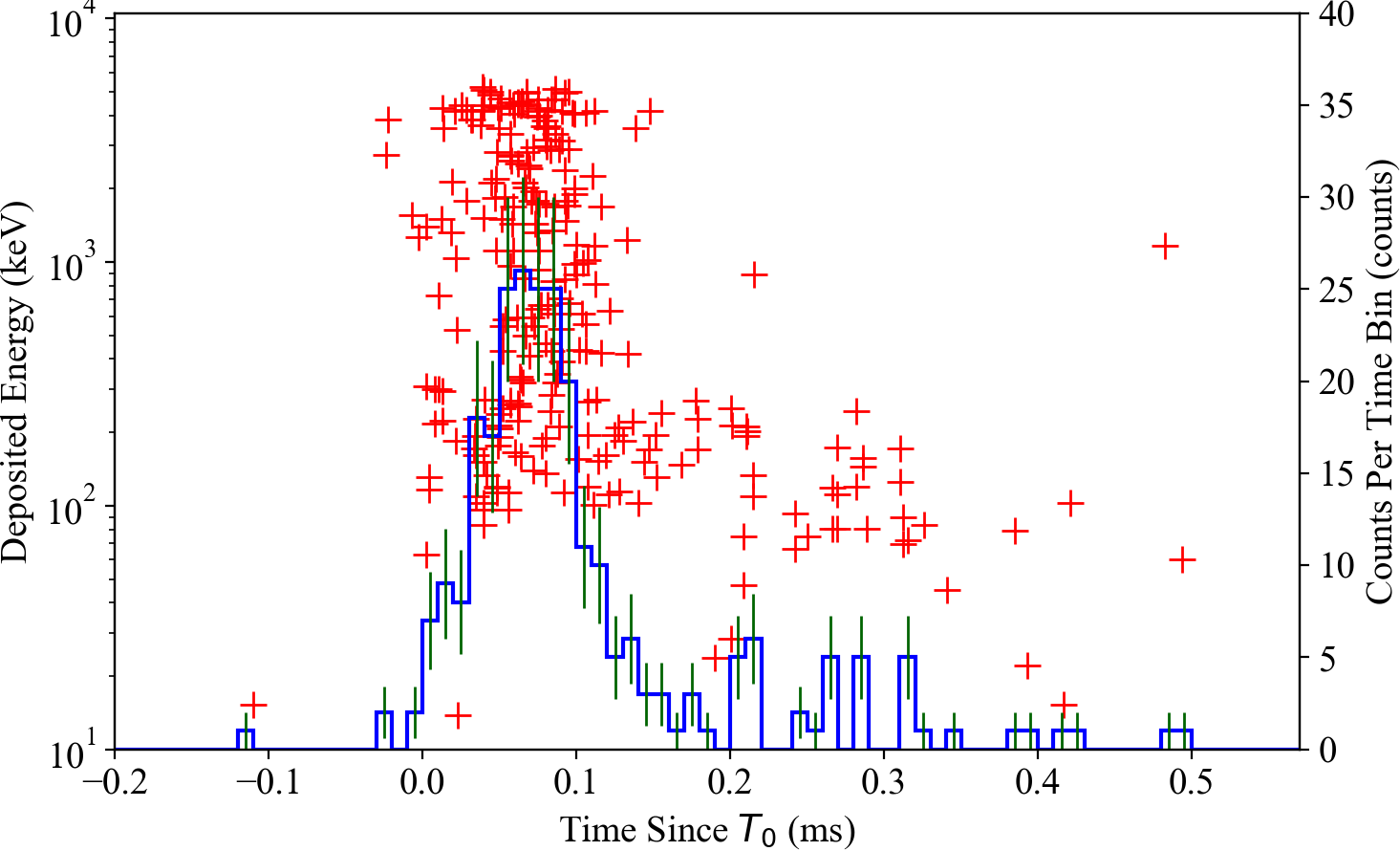}}
        \quad
        \subfigure[$T_{\rm 0}$ UT 2021-08-14T09:54:29.1772]{\includegraphics[height=3.0cm]{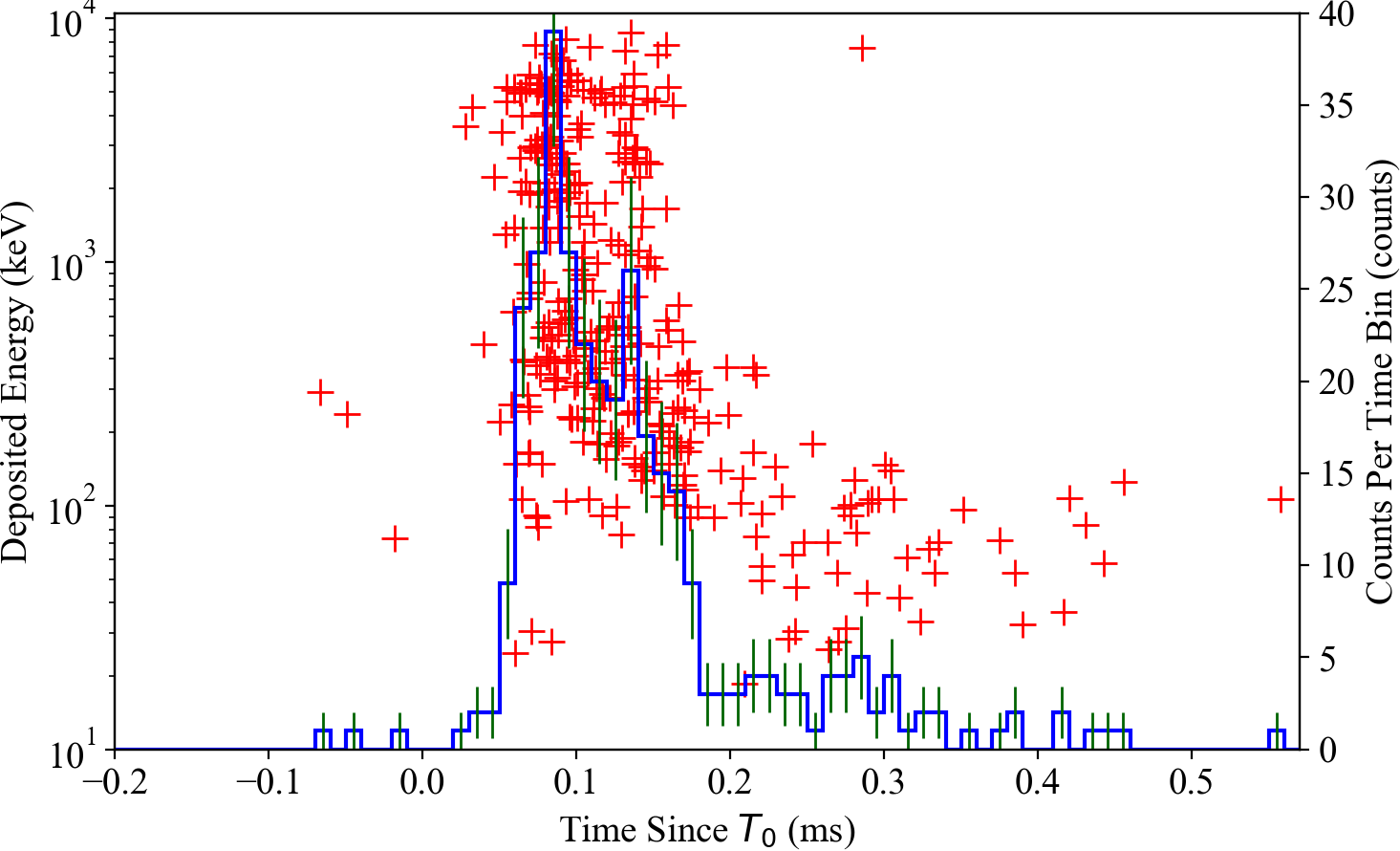}}
        \quad
        \subfigure[$T_{\rm 0}$ UT 2021-08-16T17:02:27.9080]{\includegraphics[height=3.0cm]{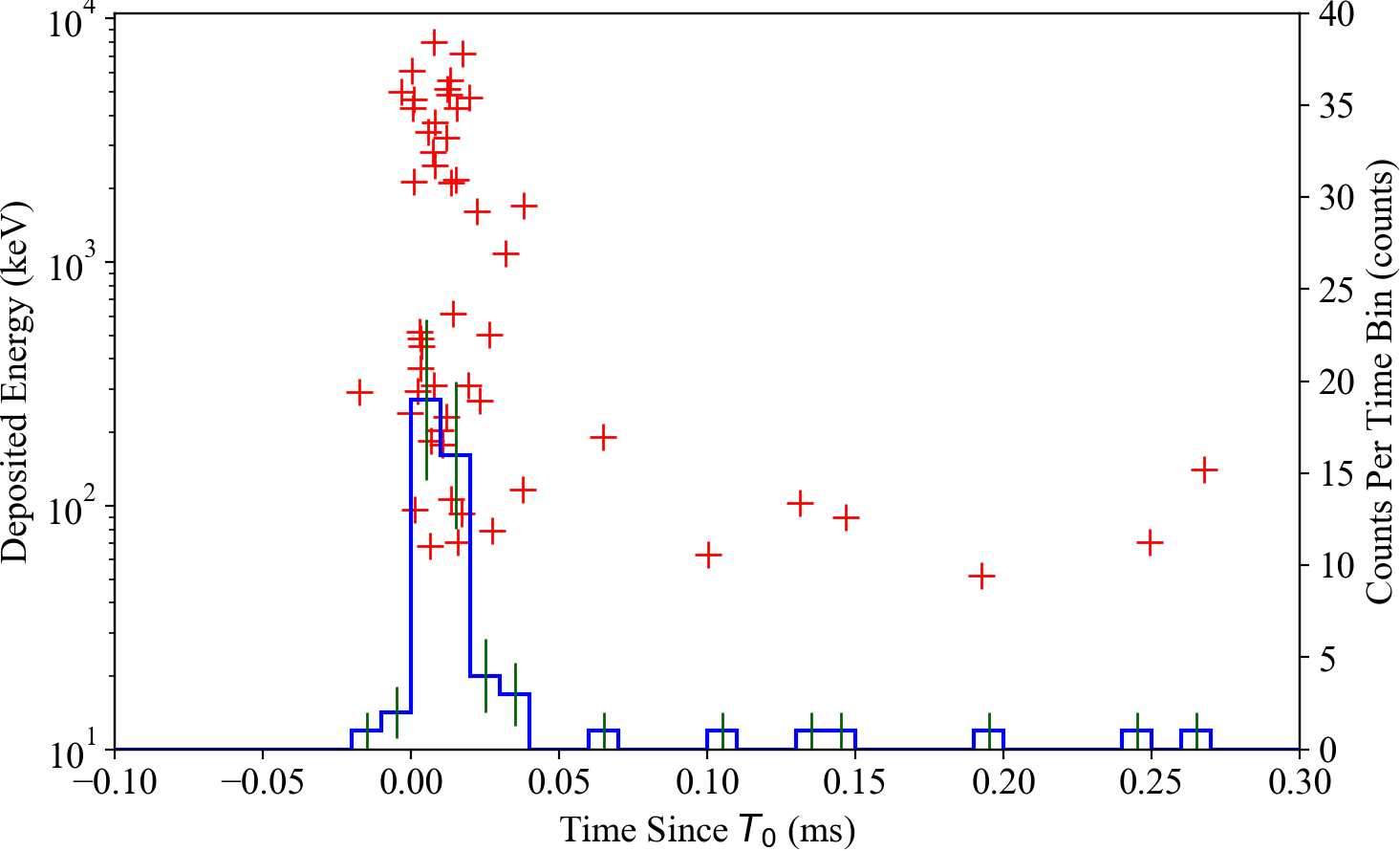}}
        \quad
        \subfigure[$T_{\rm 0}$ UT 2022-03-29T08:56:28.5994]{\includegraphics[height=3.0cm]{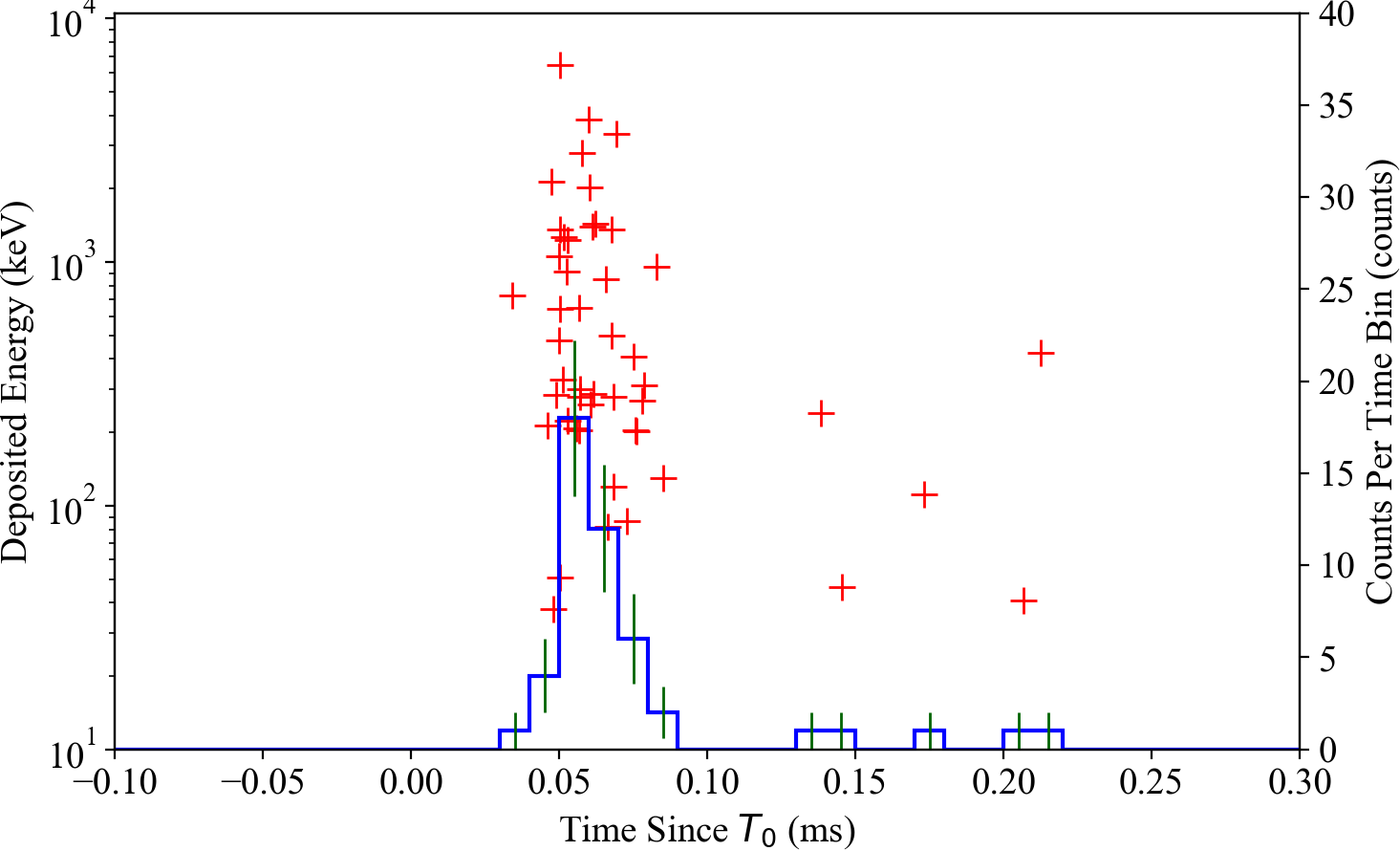}}
        \quad
        \caption{Light curves and time-energy scatters of characteristic GECAM TGFs. (a) to (c): multipeak TGFs. (d) to (f): bright TGFs with $>$150 counts in duration. (g) and (h): short-duration TGFs (20~$\mu$s and 37~$\mu$s). The black histograms and red crosses show light curves and time-energy scatters, respectively. The vertical and horizontal for all TGFs are on the same scales except for (b) and (c).}
        \label{FIG_4}
    \end{figure*}

    In Figure \ref{FIG_4}, light curves and time-energy scatter plots are illustrated for three multipeak, three bright, and two short TGFs. Note that the count clusters around $\sim$4000~keV are located at the GRDs' saturated peak. These events' recorded energy is inaccurate. This is primarily due to electronics' saturation, which leads to a signal cutoff at some stage. The pulse pile-up effect may also come into play. These effects related to the saturation peak are still under study.

    It is worth noticing an interesting double-peaked TGF (Figure \ref{FIG_4}a) that is characterized by two $\sim$100~$\mu$s pulses with very similar temporal and spectral structures. Two possible scenarios may explain this double-peaked TGF. For the first, it is accepted that the upward leader channel of a lightning discharge could branch during propagation \cite{TGF_TEO_Wu2015,TGF_TEO_Liu2022}. We speculate that such branching may reflect the complicate electric field distribution, which may result in multiple or overlapping pluses in a TGF. It could be also responsible for cases shown in Figure \ref{FIG_4}b and \ref{FIG_4}c. However, this double-peak TGF (Figure \ref{FIG_4}a) may require coincidences comparing to other TGFs in Figure \ref{FIG_4}b to \ref{FIG_4}c, i.e. two intracloud electric fields with similar distribution on the passageway of these upward leader channels. For the second, it could be associated with two successive steps of one propagating channel. We note that the time interval between the two pulses of this double-peak TGF is generally consistent with the typical duration of the stepped leader's step, i.e., $\sim$0.1~ms \cite{TGF_LED_Lyu2016}. Meanwhile, the typical length of leader steps during intracloud lightning discharge is from several hundred meters to several kilometers \cite{TGF_LIG_Stolzenburg2016}. Therefore, the second pulse of this TGF was also likely generated after the initial leader (which resulted in the first pulse) propagated forward for one or several more steps.

    The soft tail, which is caused by multiple Compton scattering of photons that makes photons arrive slightly later, is an important feature of TGFs \cite<e.g.,>{TGF_RSC_Xu2019}. The energy band of high-gain channel of GRDs could be down to $\sim$15~keV, which is efficient to charaterize these tails (see Figure \ref{FIG_4}d to \ref{FIG_4}f). 

    The existing models have shown a general correlation between gamma-ray production and intense electric field distribution \cite{TGF_TEO_Dwyer2012,TGF_TEO_Liu2013}, while these models do not fully account for the intrinsic complexity of the electric field driving mechanisms. Whether the light curve structure of TGFs detected by GECAM is related to the specific distribution of intense electric field merits further investigation with these models. Furthermore, some extreme short-duration (down to 20~$\mu$s) TGFs are found, as shown in Figure \ref{FIG_4}g to \ref{FIG_4}h.

    %

\clearpage
\section{GECAM TEBs and Two Special Events}

\begin{figure*}
        \centering
        \subfigure[TEB UT 2021-10-27T22:49:33.082]{\includegraphics[height=4.0cm]{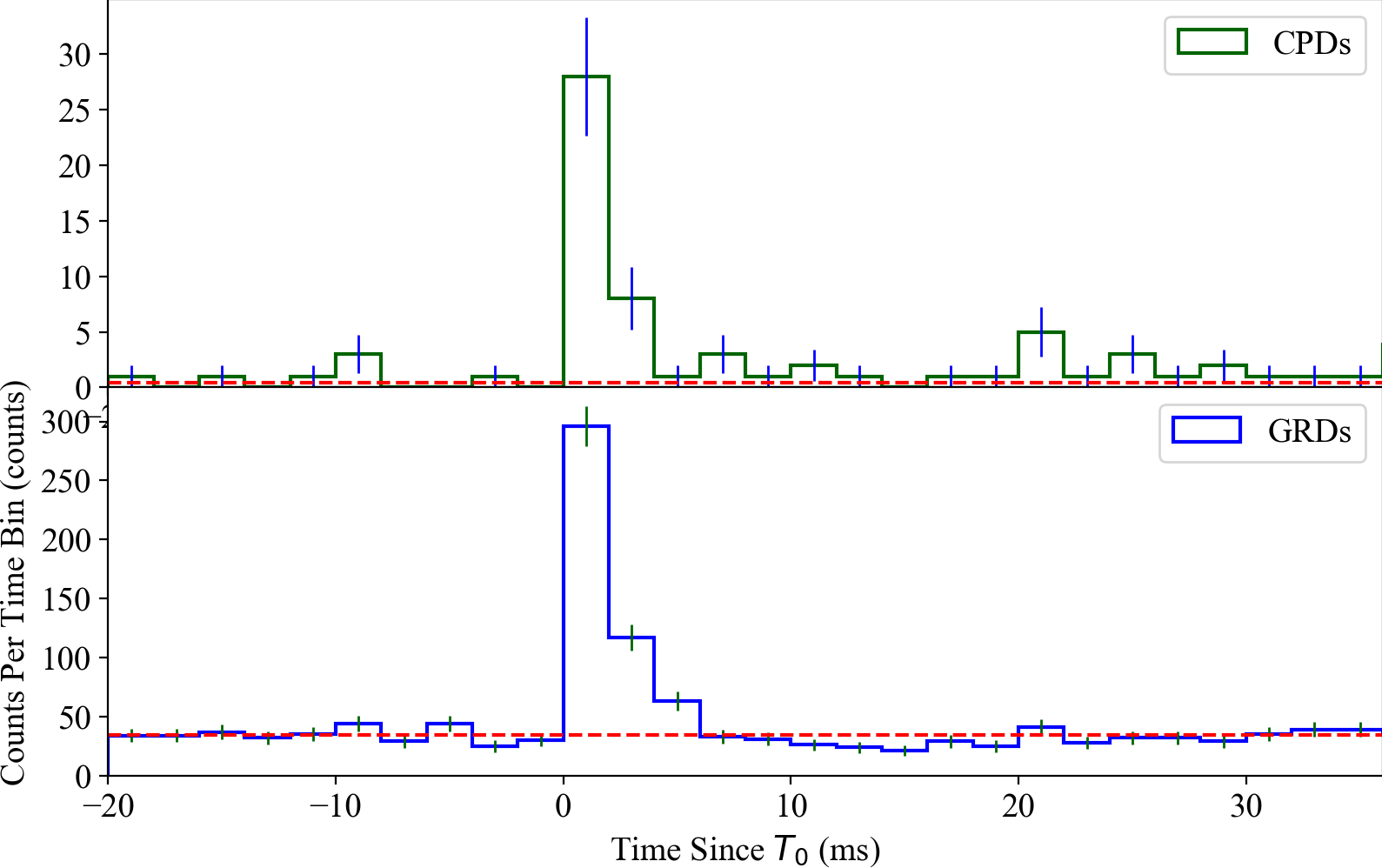}}
        \quad
        \subfigure[TEB UT 2022-07-26T00:16:13.728]{\includegraphics[height=4.0cm]{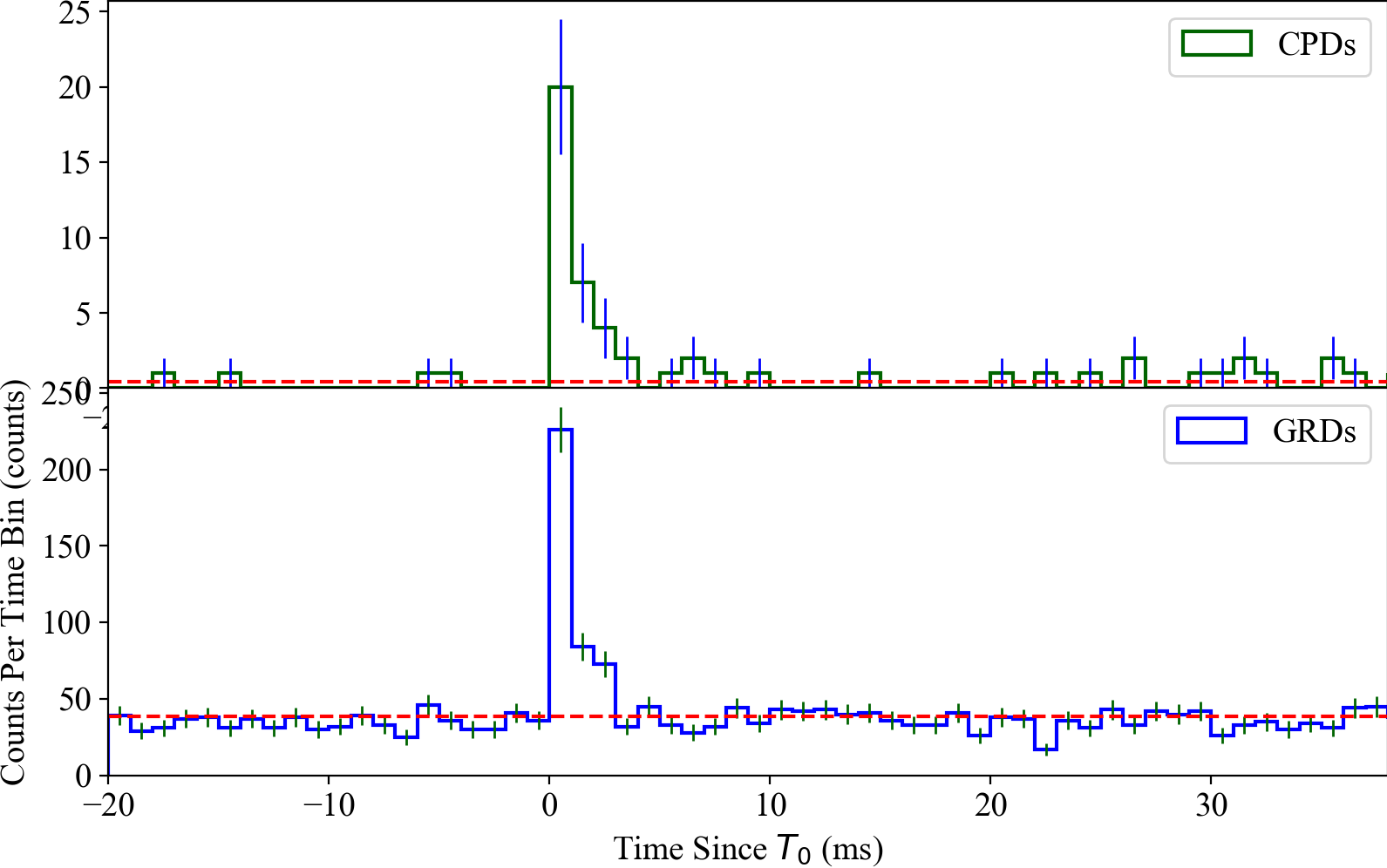}}
        \quad
        \\
        \subfigure[TEB-like event UT 2021-09-11T18:34:40.552]{\includegraphics[height=4.0cm]{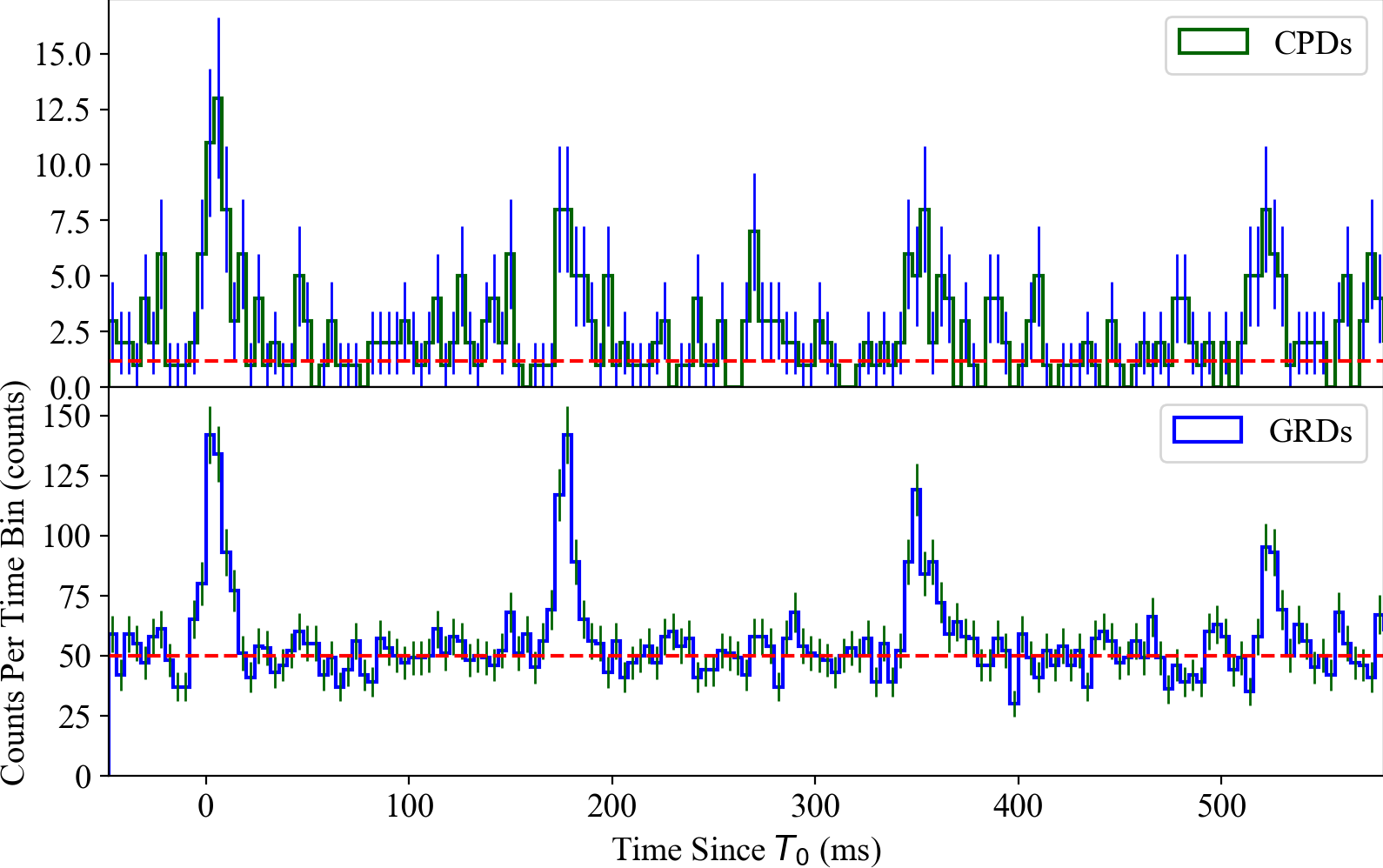}}
        \quad
        \subfigure[TEB-like event UT 2021-07-10T01:46:36.710]{\includegraphics[height=4.0cm]{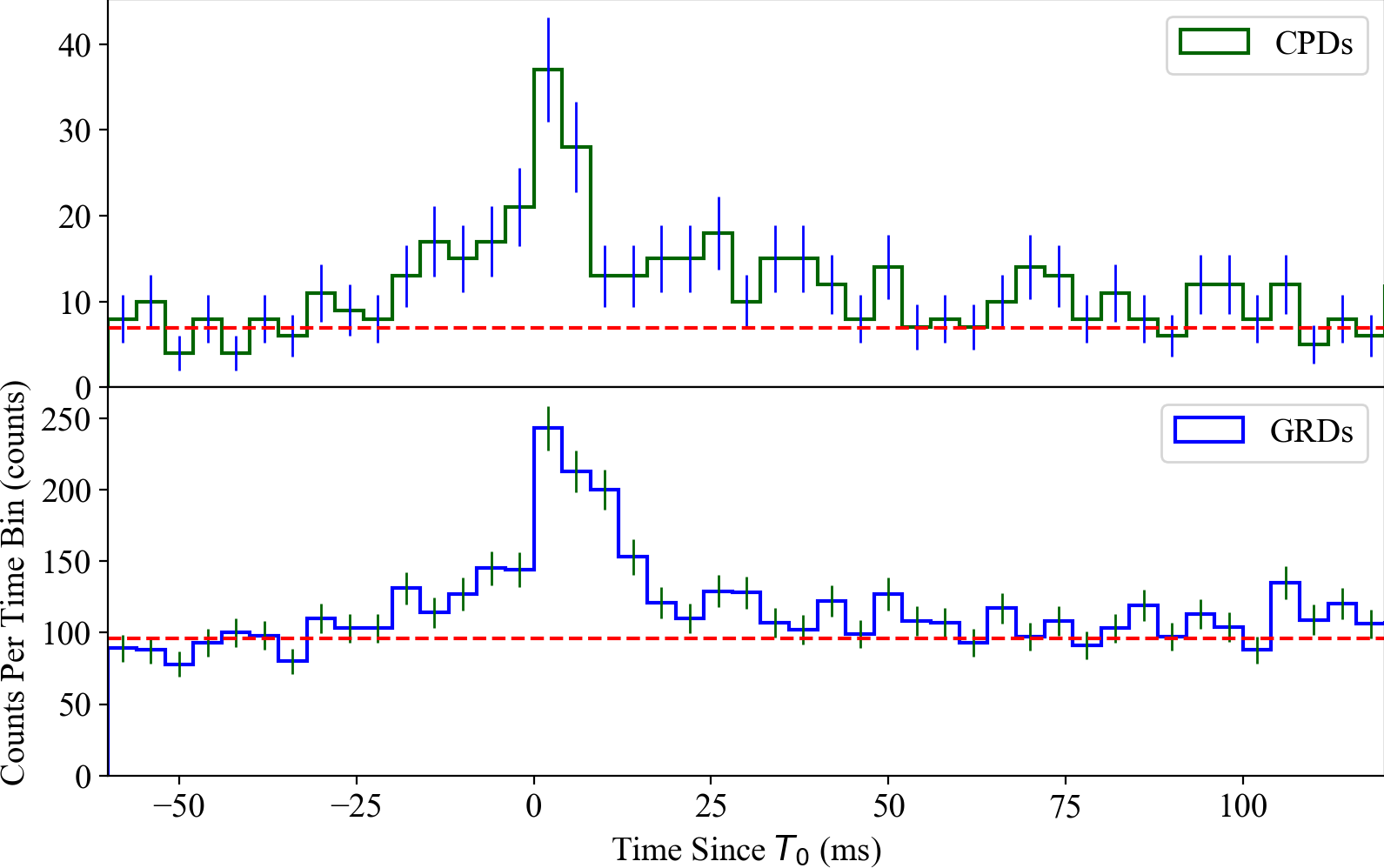}}
        \quad
        \\
        \subfigure[]{\includegraphics[height=6.5cm]{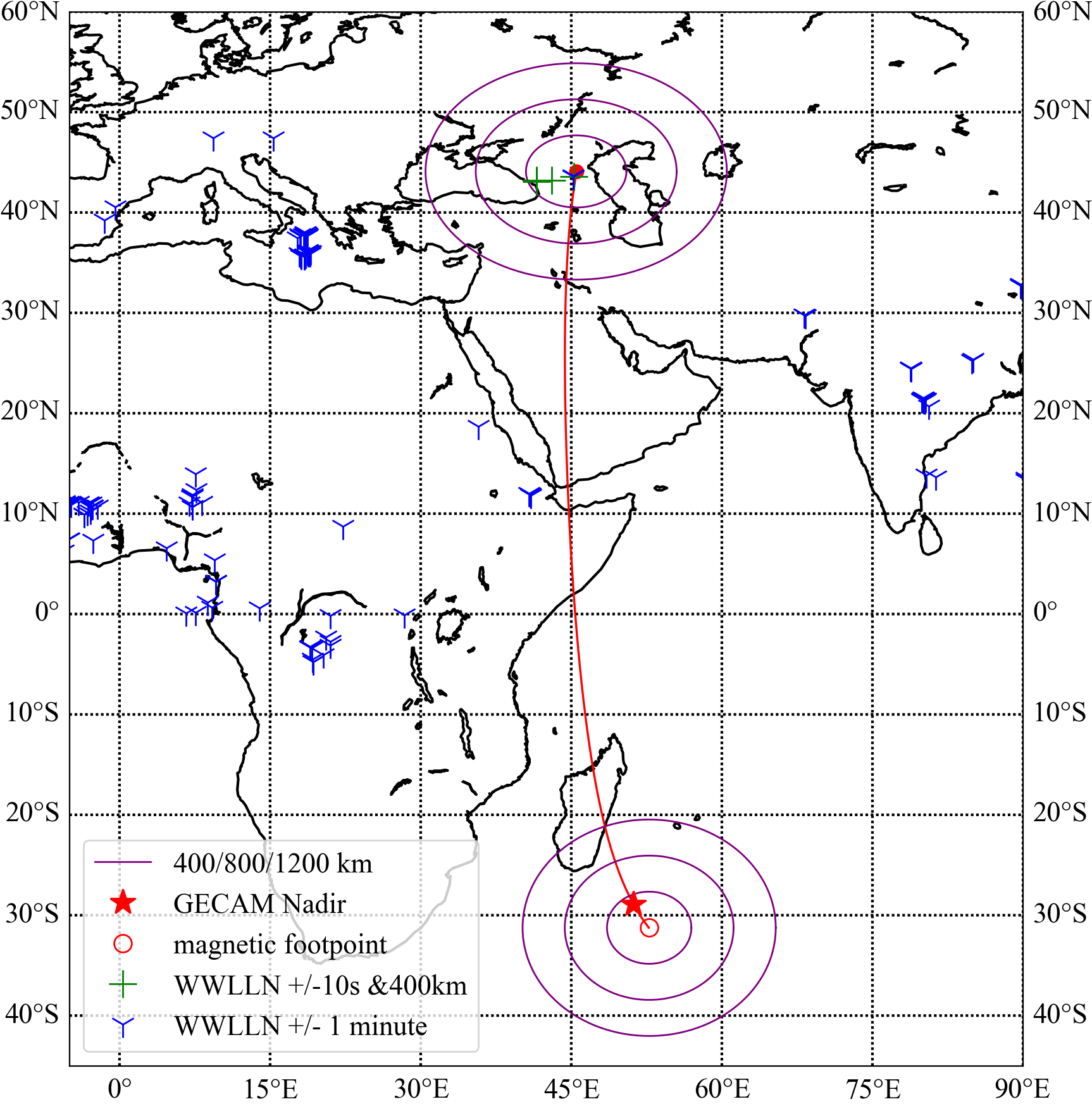}}
        \quad
        \subfigure[]{\includegraphics[height=4.5cm]{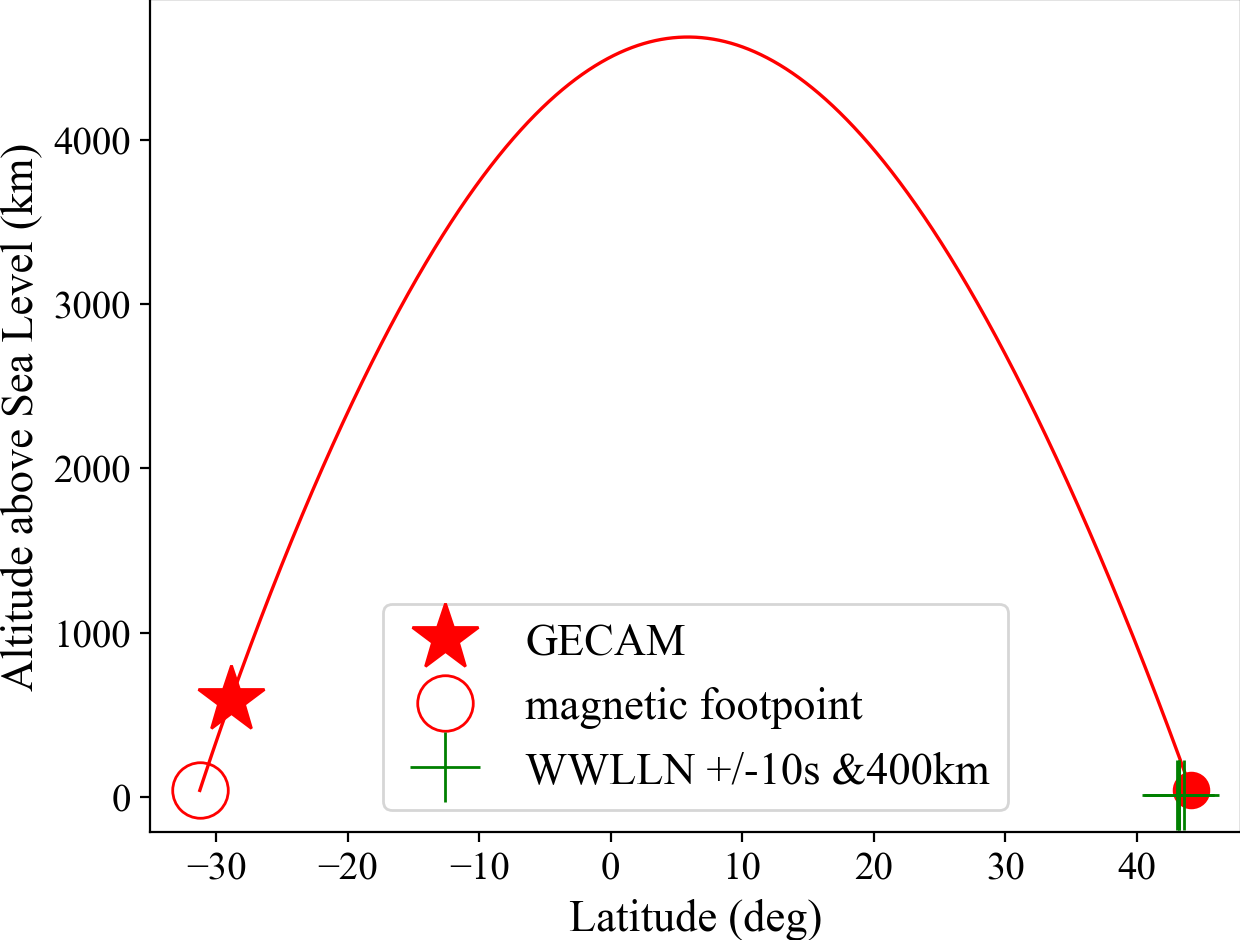}}
        \quad
        \caption{(a) to (b) Light curves of two high-confidence TEBs. (c) to (d) Light curves of two special TEB-like events. For each event, the upper and lower panels show light curves of CPDs and GRDs, respectively. (e) Map of GECAM nadir (red star), WWLLN lightning (blue triangles and green pluses), the traced magnetic field line (red line), and their footpoints (red circles) for the event shown in subfigure (c). (f) The latitude-altitude projected map of the event shown in subfigure (c). The blue triangles illustrate total WWLLN detections within $\pm$60~seconds, and green pluses illustrate the WWLLN lightning around the northern magnetic footpoint 400~km within 10~seconds. The solid red circle shows the northern magnetic footpoint and the hollow red circle shows the southern magnetic footpoint.}
        \label{FIG_5}
    \end{figure*}

    Here, we first present two high-confidence TEBs, as shown in Figure \ref{FIG_2}a, Figure \ref{FIG_3}d, Figure \ref{FIG_5}a and \ref{FIG_5}b. GECAM CPDs are mostly used to detect electrons and positrons in orbit, since it has low detection efficiency to gamma-ray \cite{GEC_INS_Xv2021}. Although TEBs can also produce many counts in GRDs as TGFs, their duration and CPD/GRD counts ratio are remarkably different from TGFs. To distinguish between TGFs and TEBs, we find a very effective criteria considering the duration and CPD/GRD counts ratio (see Figure \ref{FIG_3}d). It is explicitly shown in Figure \ref{FIG_3}d that TEBs and TGFs are separated into two groups according to duration and CPD/GRD counts ratio. Note that the negative values of the CPD/GRD counts ratio mean no significant excess counts registered in CPDs. The duration of TGFs ($<$1~ms) and TEBs ($>$2~ms) are also distinctively different.

    In addition to these two high-confidence TEBs above, GECAM-B also detected two special events (see Figure \ref{FIG_5}c and \ref{FIG_5}d). Based on the criteria presented in Figure \ref{FIG_3}d, they could be classified as "TEBs". However, their slow-rise light curves deviate from the characteristics of previously reported TEBs \cite{TEB_BAT_Dwyer2008, TGF_GBM_Roberts2018, TEB_ASM_Sarria2019, TGF_AGL_Lindanger2020}, although the third and fourth pulse of Figure \ref{FIG_5}c seem to display a fast-rise light curve of typical TEBs.

    Particularly, the special event in Figure \ref{FIG_5}c consists of quadruple pulses and was detected by GECAM-B over the Southwest Indian Ocean at 18:34:40.551997 UTC on September 11th, 2021. Following previous TEB studies, we trace the geomagnetic line using the International Geomagnetic Reference Field (IGRF-13) model \cite{TEB_IGRF13_Alken2021}, since TEB electrons and positrons will travel along the Earth's magnetic field lines. There is no lightning activity around the GECAM-B nadir (51.2$^{\circ}$~E, 28.9$^{\circ}$~S, 587.8~km) and the southern magnetic footpoint (52.8$^{\circ}$~E, 31.3$^{\circ}$~S, 40~km, 37379~nT) within $\pm$1~minute and a radius of 1200~km (see Figure \ref{FIG_5}e). The GECAM-B is relatively close to the southern magnetic footpoint, with a magnetic line path length of $\sim$600~km (see Figure \ref{FIG_5}f). But, there is a cluster of WWLLN lightning around the northern magnetic footpoint (44.1$^{\circ}$~E, 45.5$^{\circ}$~N, 40~km, 50129~nT) within $\pm$10~seconds and a radius of 400~km. Therefore, we think that the electrons and positrons should originate from the vicinity of the northern footpoint. We note that the magnitude of geomagnetic field given by the IGRF-13 model at the northern footpoint (at 40~km altitude) is higher than that of the southern footpoint, thus there should be not return peak for this event.

    The time intervals between each neighboring pulse in the quad-peaked event are comparable, i.e., $\sim$169~ms, $\sim$175~ms, and $\sim$172~ms, respectively. We note that there are cases in the \textit{Fermi}/GBM TGF sample where the time interval between two TGFs is approximately hundreds of milliseconds \cite{TGF_GBM_Roberts2018}. It is possible that there are quadruple or more neighboring TGFs with similar time intervals. Indeed, while examining the lightning dataset for other time than this special TEB-like event, we find that there are some lightning processes consisting of four lightning strokes with waiting time of $\sim$160~ms to $\sim$180~ms. These lightning strokes either originate from the same location (within location error) or from within a small region of $\sim$30~km. We speculate that the quad-peaked event may be produced by such kind of lightning process around the northern footpoint. If this TEB-like event is from four TGFs, they should have some connections, e.g. the periodic TGFs \cite{TGF_ASM_Kochkin2019, TGF_ASM_Ostgaard2019}, and the distance between these four TGFs should be not very far, otherwise they would not be detected as a single TEB-like event by GECAM-B. Besides, the production and propagation mechanisms of this TEB require more investigation to explain the atypical light curve.

    It is also possible that it represents a new, unidentified class of event. Therefore, based on our current knowledge, we classify these two events as special TEB-like events. Detailed analysis will be reported in a forthcoming work.

    %

\clearpage
\section{Conclusion}

    With novel designs on detectors and electronics, GECAM is a new powerful instrument to detect and identify TGFs and TEBs, as well as study their temporal and spectral properties. Thanks to the high time resolution (100~ns), broad detection energy range ($\sim$15~keV to $\sim$5~MeV) and anti-data-saturation designs, GECAM can record very bright TGFs and TEBs, and reveal their fine structures in light curves and spectrum, which can help us better understand the production mechanism of TGFs and TEBs.

    In this paper, a dedicated search algorithm of TGF and TEB has been implemented for GECAM, which results in 147 bright TGFs, 2 typical TEBs and 2 special TEB-like events in $\sim$9 months of data. TGF detection rate for GECAM-B is $\sim$200 TGFs/year, which will increase if we loose the search threshold. A very high TGF-lightning association rate of $\sim$80\% is obtained between GECAM and GLD360 in east Asia region. Some interesting TGFs are found, such as a double-peak TGF with very similar temporal and spectral distribution.

    For most gamma-ray space telescopes, disentangling TEBs usually rely on the 511~keV line feature in the spectrum or the return peak in light curve. With joint observation of GRDs and CPDs, GECAM can distinguish between TGFs and TEBs according to the duration distribution and CPD/GRD counts ratio.

    Interestingly, GECAM discovered two special TEB-like events, and one of them has quadruple peaks which probably originated from a special lightning discharge process. The nature of these TEB-like events remains to be revealed which requires a dedicated in-depth study. This kind of events may shed new light on the TGF and TEB mechanism.

\section*{Open Research}
    All data that are used to produce the figures in this paper have been uploaded to Zenodo with DOI: 10.5281/zenodo.8028217 (\url{https://zenodo.org/record/8028217}) \cite{YZ_TGF_GEC_SupDat_03}, available under Creative Commons Attribution 4.0 International License. The World Wide Lightning Location Network (WWLLN) and GLD360 data used in this paper are also available from the Zenodo repository. The authors wish to thank the WWLLN, a collaboration among over 50 universities and institutions, for providing the lightning location data used in these datasets and in the paper. Additional WWLLN data are available at a nominal cost from \url{http://wwlln.net}. Researchers may contact Vaisala at \url{https://www.vaisala.com/en/lp/contact-us-lightningsolutions} to arrange research use of additional GLD360 data \cite{YZ_TGF_GEC_SupDat_02}.

\acknowledgments
    The GECAM (HuaiRou-1) mission is supported by Strategic Priority Research Program on Space Science of Chinese Academy of Sciences, China. We thank the support from National Key R\&D Program of China (2021YFA0718500), Strategic Priority Research Program on Space Science, Chinese Academy of Sciences (Grant No. XDA15360102, XDA15360300, XDA15052700), National Natural Science Foundation of China (Grant No. 12273042, 12173038, 42274205, U1938115, U2038106), National HEP Data Center (Grant No. E029S2S1) and the open fund of Hubei Luojia Laboratory (Grant No. 220100051). We thank Xi Long (Harvard University) for helpful discussions. GLD360 data used in this paper belong to Vaisala Inc who supports the ASIM project. The authors wish to thank the World Wide Lightning Location Network (\url{http://wwlln.net}) as a collaboration of more than 50 universities.

\clearpage

\begin{thebibliography}{}

    \bibitem [\protect \citeauthoryear {%
    Alken%
    \ \protect \BOthers {.}}{%
    Alken%
    \ \protect \BOthers {.}}{%
    {\protect \APACyear {2021}}%
    }]{%
    TEB_IGRF13_Alken2021}
    \APACinsertmetastar {%
    TEB_IGRF13_Alken2021}%
    \begin{APACrefauthors}%
    Alken, P.%
    , Th{\'e}bault, E.%
    , Beggan, C\BPBI D.%
    , Amit, H.%
    , Aubert, J.%
    , Baerenzung, J.%
    \BDBL {}others%
    \end{APACrefauthors}%
    \unskip\
    \newblock
    \APACrefYearMonthDay{2021}{}{}.
    \newblock
    {\BBOQ}\APACrefatitle {International geomagnetic reference field: the
      thirteenth generation} {International geomagnetic reference field: the
      thirteenth generation}.{\BBCQ}
    \newblock
    \APACjournalVolNumPages{Earth, Planets and Space}{73}{1}{1--25}.
    \PrintBackRefs{\CurrentBib}

    \bibitem [\protect \citeauthoryear {%
    An%
    \ \protect \BOthers {.}}{%
    An%
    \ \protect \BOthers {.}}{%
    {\protect \APACyear {2022}}%
    }]{%
    GEC_INS_An2022}
    \APACinsertmetastar {%
    GEC_INS_An2022}%
    \begin{APACrefauthors}%
    An, Z\BPBI H.%
    , Sun, X\BPBI L.%
    , Zhang, D\BPBI L.%
    , Yang, S.%
    , Li, X\BPBI Q.%
    , Wen, X\BPBI Y.%
    \BDBL {}Zhou, X.%
    \end{APACrefauthors}%
    \unskip\
    \newblock
    \APACrefYearMonthDay{2022}{}{}.
    \newblock
    {\BBOQ}\APACrefatitle {The design and performance of GRD onboard the GECAM
      satellite} {The design and performance of grd onboard the gecam
      satellite}{\BBCQ}\ [Journal Article].
    \newblock
    \APACjournalVolNumPages{Radiation Detection Technology and
      Methods}{6}{1}{43-52}.
    \newblock
    \begin{APACrefURL} \url{https://doi.org/10.1007/s41605-021-00289-y}
      \end{APACrefURL}
    \newblock
    \begin{APACrefDOI} \doi{10.1007/s41605-021-00289-y} \end{APACrefDOI}
    \PrintBackRefs{\CurrentBib}

    \bibitem [\protect \citeauthoryear {%
    {Belz}%
    \ \protect \BOthers {.}}{%
    {Belz}%
    \ \protect \BOthers {.}}{%
    {\protect \APACyear {2020}}%
    }]{%
    TGF_GND_Belz2020}
    \APACinsertmetastar {%
    TGF_GND_Belz2020}%
    \begin{APACrefauthors}%
    {Belz}, J\BPBI W.%
    , {Krehbiel}, P\BPBI R.%
    , {Remington}, J.%
    , {Stanley}, M\BPBI A.%
    , {Abbasi}, R\BPBI U.%
    , {LeVon}, R.%
    \BDBL {}{Zundel}, Z.%
    \end{APACrefauthors}%
    \unskip\
    \newblock
    \APACrefYearMonthDay{2020}{{\APACmonth{12}}}{}.
    \newblock
    {\BBOQ}\APACrefatitle {{Observations of the Origin of Downward Terrestrial
      Gamma-Ray Flashes}} {{Observations of the Origin of Downward Terrestrial
      Gamma-Ray Flashes}}.{\BBCQ}
    \newblock
    \APACjournalVolNumPages{Journal of Geophysical Research:
      Atmospheres}{125}{23}{e31940}.
    \newblock
    \begin{APACrefDOI} \doi{10.1029/2019JD031940} \end{APACrefDOI}
    \PrintBackRefs{\CurrentBib}

    \bibitem [\protect \citeauthoryear {%
    {Bhat}%
    \ \protect \BOthers {.}}{%
    {Bhat}%
    \ \protect \BOthers {.}}{%
    {\protect \APACyear {2014}}%
    }]{%
    INS_EFF_Bhat2014}
    \APACinsertmetastar {%
    INS_EFF_Bhat2014}%
    \begin{APACrefauthors}%
    {Bhat}, P\BPBI N.%
    , {Fishman}, G\BPBI J.%
    , {Briggs}, M\BPBI S.%
    , {Connaughton}, V.%
    , {Meegan}, C\BPBI A.%
    , {Paciesas}, W\BPBI S.%
    \BDBL {}{Xiong}, S.%
    \end{APACrefauthors}%
    \unskip\
    \newblock
    \APACrefYearMonthDay{2014}{{\APACmonth{11}}}{}.
    \newblock
    {\BBOQ}\APACrefatitle {{Fermi gamma-ray burst monitor detector performance at
      very high counting rates}} {{Fermi gamma-ray burst monitor detector
      performance at very high counting rates}}.{\BBCQ}
    \newblock
    \APACjournalVolNumPages{Experimental Astronomy}{38}{1-2}{331-357}.
    \newblock
    \begin{APACrefDOI} \doi{10.1007/s10686-014-9424-z} \end{APACrefDOI}
    \PrintBackRefs{\CurrentBib}

    \bibitem [\protect \citeauthoryear {%
    {Briggs}%
    \ \protect \BOthers {.}}{%
    {Briggs}%
    \ \protect \BOthers {.}}{%
    {\protect \APACyear {2010}}%
    }]{%
    TGF_GBM_Briggs2010}
    \APACinsertmetastar {%
    TGF_GBM_Briggs2010}%
    \begin{APACrefauthors}%
    {Briggs}, M\BPBI S.%
    , Fishman, G.%
    , Connaughton, V.%
    , Bhat, P.%
    , Paciesas, W.%
    , Preece, R.%
    \BDBL {}others%
    \end{APACrefauthors}%
    \unskip\
    \newblock
    \APACrefYearMonthDay{2010}{}{}.
    \newblock
    {\BBOQ}\APACrefatitle {First results on terrestrial gamma ray flashes from the
      Fermi Gamma-ray Burst Monitor} {First results on terrestrial gamma ray
      flashes from the fermi gamma-ray burst monitor}.{\BBCQ}
    \newblock
    \APACjournalVolNumPages{Journal of Geophysical Research: Space
      Physics}{115}{A7}{}.
    \PrintBackRefs{\CurrentBib}

    \bibitem [\protect \citeauthoryear {%
    {Briggs}%
    \ \protect \BOthers {.}}{%
    {Briggs}%
    \ \protect \BOthers {.}}{%
    {\protect \APACyear {2013}}%
    }]{%
    TGF_GBM_Briggs2013}
    \APACinsertmetastar {%
    TGF_GBM_Briggs2013}%
    \begin{APACrefauthors}%
    {Briggs}, M\BPBI S.%
    , Xiong, S.%
    , Connaughton, V.%
    , Tierney, D.%
    , Fitzpatrick, G.%
    , Foley, S.%
    \BDBL {}others%
    \end{APACrefauthors}%
    \unskip\
    \newblock
    \APACrefYearMonthDay{2013}{}{}.
    \newblock
    {\BBOQ}\APACrefatitle {Terrestrial gamma-ray flashes in the Fermi era: Improved
      observations and analysis methods} {Terrestrial gamma-ray flashes in the
      fermi era: Improved observations and analysis methods}.{\BBCQ}
    \newblock
    \APACjournalVolNumPages{Journal of Geophysical Research: Space
      Physics}{118}{6}{3805--3830}.
    \PrintBackRefs{\CurrentBib}

    \bibitem [\protect \citeauthoryear {%
    Cai%
    \ \protect \BOthers {.}}{%
    Cai%
    \ \protect \BOthers {.}}{%
    {\protect \APACyear {2021}}%
    }]{%
    HXM_SEA_Cai2021}
    \APACinsertmetastar {%
    HXM_SEA_Cai2021}%
    \begin{APACrefauthors}%
    Cai, C.%
    , Xiong, S\BPBI L.%
    , Li, C\BPBI K.%
    , Liu, C\BPBI Z.%
    , Zhang, S\BPBI N.%
    , Li, X\BPBI B.%
    \BDBL {}Zhou, D\BPBI K.%
    \end{APACrefauthors}%
    \unskip\
    \newblock
    \APACrefYearMonthDay{2021}{09}{}.
    \newblock
    {\BBOQ}\APACrefatitle {{Search for gamma-ray bursts and gravitational wave
      electromagnetic counterparts with High Energy X-ray Telescope of
      Insight-HXMT}} {{Search for gamma-ray bursts and gravitational wave
      electromagnetic counterparts with High Energy X-ray Telescope of
      Insight-HXMT}}.{\BBCQ}
    \newblock
    \APACjournalVolNumPages{Monthly Notices of the Royal Astronomical
      Society}{508}{3}{3910-3920}.
    \newblock
    \begin{APACrefURL} \url{https://doi.org/10.1093/mnras/stab2760}
      \end{APACrefURL}
    \newblock
    \begin{APACrefDOI} \doi{10.1093/mnras/stab2760} \end{APACrefDOI}
    \PrintBackRefs{\CurrentBib}

    \bibitem [\protect \citeauthoryear {%
    Cai%
    \ \protect \BOthers {.}}{%
    Cai%
    \ \protect \BOthers {.}}{%
    {\protect \APACyear {2022}}%
    }]{%
    HXM_SEA_Cai2023}
    \APACinsertmetastar {%
    HXM_SEA_Cai2023}%
    \begin{APACrefauthors}%
    Cai, C.%
    , Xiong, S\BHBI L.%
    , Xue, W\BHBI C.%
    , Zhao, Y.%
    , Xiao, S.%
    , Yi, Q\BHBI B.%
    \BDBL {}Zhang, F.%
    \end{APACrefauthors}%
    \unskip\
    \newblock
    \APACrefYearMonthDay{2022}{10}{}.
    \newblock
    {\BBOQ}\APACrefatitle {{Burst search method based on likelihood ratio in
      Poisson statistics}} {{Burst search method based on likelihood ratio in
      Poisson statistics}}.{\BBCQ}
    \newblock
    \APACjournalVolNumPages{Monthly Notices of the Royal Astronomical
      Society}{518}{2}{2005-2014}.
    \newblock
    \begin{APACrefURL} \url{https://doi.org/10.1093/mnras/stac3075}
      \end{APACrefURL}
    \newblock
    \begin{APACrefDOI} \doi{10.1093/mnras/stac3075} \end{APACrefDOI}
    \PrintBackRefs{\CurrentBib}

    \bibitem [\protect \citeauthoryear {%
    Celestin%
    \ \BBA {} Pasko%
    }{%
    Celestin%
    \ \BBA {} Pasko%
    }{%
    {\protect \APACyear {2011}}%
    }]{%
    TGF_TEO_Celestin2011}
    \APACinsertmetastar {%
    TGF_TEO_Celestin2011}%
    \begin{APACrefauthors}%
    Celestin, S.%
    \BCBT {}\ \BBA {} Pasko, V\BPBI P.%
    \end{APACrefauthors}%
    \unskip\
    \newblock
    \APACrefYearMonthDay{2011}{}{}.
    \newblock
    {\BBOQ}\APACrefatitle {Energy and fluxes of thermal runaway electrons produced
      by exponential growth of streamers during the stepping of lightning leaders
      and in transient luminous events} {Energy and fluxes of thermal runaway
      electrons produced by exponential growth of streamers during the stepping of
      lightning leaders and in transient luminous events}.{\BBCQ}
    \newblock
    \APACjournalVolNumPages{Journal of Geophysical Research: Space
      Physics}{116}{A3}{}.
    \PrintBackRefs{\CurrentBib}

    \bibitem [\protect \citeauthoryear {%
    Celestin%
    , Xu%
    \BCBL {}\ \BBA {} Pasko%
    }{%
    Celestin%
    \ \protect \BOthers {.}}{%
    {\protect \APACyear {2013}}%
    }]{%
    TGF_LIG_Celestin2013}
    \APACinsertmetastar {%
    TGF_LIG_Celestin2013}%
    \begin{APACrefauthors}%
    Celestin, S.%
    , Xu, W.%
    \BCBL {}\ \BBA {} Pasko, V.%
    \end{APACrefauthors}%
    \unskip\
    \newblock
    \APACrefYearMonthDay{2013}{}{}.
    \newblock
    {\BBOQ}\APACrefatitle {Spectra of X-ray and Gamma-ray Bursts Produced by
      Stepping Lightning Leaders} {Spectra of x-ray and gamma-ray bursts produced
      by stepping lightning leaders}.{\BBCQ}
    \newblock
    \BIn{} \APACrefbtitle {EGU General Assembly Conference Abstracts} {Egu general
      assembly conference abstracts}\ (\BPG~13065).
    \PrintBackRefs{\CurrentBib}

    \bibitem [\protect \citeauthoryear {%
    Chanrion%
    \ \BBA {} Neubert%
    }{%
    Chanrion%
    \ \BBA {} Neubert%
    }{%
    {\protect \APACyear {2010}}%
    }]{%
    TGF_TEO_Chanrion2010}
    \APACinsertmetastar {%
    TGF_TEO_Chanrion2010}%
    \begin{APACrefauthors}%
    Chanrion, O.%
    \BCBT {}\ \BBA {} Neubert, T.%
    \end{APACrefauthors}%
    \unskip\
    \newblock
    \APACrefYearMonthDay{2010}{}{}.
    \newblock
    {\BBOQ}\APACrefatitle {Production of runaway electrons by negative streamer
      discharges} {Production of runaway electrons by negative streamer
      discharges}.{\BBCQ}
    \newblock
    \APACjournalVolNumPages{Journal of Geophysical Research: Space
      Physics}{115}{A6}{}.
    \PrintBackRefs{\CurrentBib}

    \bibitem [\protect \citeauthoryear {%
    Dwyer%
    }{%
    Dwyer%
    }{%
    {\protect \APACyear {2003}}%
    }]{%
    TGF_TEO_Dwyer2003}
    \APACinsertmetastar {%
    TGF_TEO_Dwyer2003}%
    \begin{APACrefauthors}%
    Dwyer, J.%
    \end{APACrefauthors}%
    \unskip\
    \newblock
    \APACrefYearMonthDay{2003}{}{}.
    \newblock
    {\BBOQ}\APACrefatitle {A fundamental limit on electric fields in air} {A
      fundamental limit on electric fields in air}.{\BBCQ}
    \newblock
    \APACjournalVolNumPages{Geophysical Research Letters}{30}{20}{}.
    \PrintBackRefs{\CurrentBib}

    \bibitem [\protect \citeauthoryear {%
    {Dwyer}%
    }{%
    {Dwyer}%
    }{%
    {\protect \APACyear {2008}}%
    }]{%
    TGF_TEO_Dwyer2008}
    \APACinsertmetastar {%
    TGF_TEO_Dwyer2008}%
    \begin{APACrefauthors}%
    {Dwyer}, J.%
    \end{APACrefauthors}%
    \unskip\
    \newblock
    \APACrefYearMonthDay{2008}{}{}.
    \newblock
    {\BBOQ}\APACrefatitle {Source mechanisms of terrestrial gamma-ray flashes}
      {Source mechanisms of terrestrial gamma-ray flashes}.{\BBCQ}
    \newblock
    \APACjournalVolNumPages{Journal of Geophysical Research:
      Atmospheres}{113}{D10}{}.
    \PrintBackRefs{\CurrentBib}

    \bibitem [\protect \citeauthoryear {%
    {Dwyer}%
    }{%
    {Dwyer}%
    }{%
    {\protect \APACyear {2012}}%
    }]{%
    TGF_TEO_Dwyer2012}
    \APACinsertmetastar {%
    TGF_TEO_Dwyer2012}%
    \begin{APACrefauthors}%
    {Dwyer}, J.%
    \end{APACrefauthors}%
    \unskip\
    \newblock
    \APACrefYearMonthDay{2012}{}{}.
    \newblock
    {\BBOQ}\APACrefatitle {The relativistic feedback discharge model of terrestrial
      gamma ray flashes} {The relativistic feedback discharge model of terrestrial
      gamma ray flashes}.{\BBCQ}
    \newblock
    \APACjournalVolNumPages{Journal of Geophysical Research: Space
      Physics}{117}{A2}{}.
    \PrintBackRefs{\CurrentBib}

    \bibitem [\protect \citeauthoryear {%
    Dwyer%
    }{%
    Dwyer%
    }{%
    {\protect \APACyear {2010}}%
    }]{%
    TGF_LIG_Dwyer2010}
    \APACinsertmetastar {%
    TGF_LIG_Dwyer2010}%
    \begin{APACrefauthors}%
    Dwyer, J\BPBI R.%
    \end{APACrefauthors}%
    \unskip\
    \newblock
    \APACrefYearMonthDay{2010}{}{}.
    \newblock
    {\BBOQ}\APACrefatitle {Diffusion of relativistic runaway electrons and
      implications for lightning initiation} {Diffusion of relativistic runaway
      electrons and implications for lightning initiation}.{\BBCQ}
    \newblock
    \APACjournalVolNumPages{Journal of Geophysical Research: Space
      Physics}{115}{A3}{}.
    \PrintBackRefs{\CurrentBib}

    \bibitem [\protect \citeauthoryear {%
    {Dwyer}%
    , Grefenstette%
    \BCBL {}\ \BBA {} Smith%
    }{%
    {Dwyer}%
    \ \protect \BOthers {.}}{%
    {\protect \APACyear {2008}}%
    }]{%
    TEB_BAT_Dwyer2008}
    \APACinsertmetastar {%
    TEB_BAT_Dwyer2008}%
    \begin{APACrefauthors}%
    {Dwyer}, J\BPBI R.%
    , Grefenstette, B\BPBI W.%
    \BCBL {}\ \BBA {} Smith, D\BPBI M.%
    \end{APACrefauthors}%
    \unskip\
    \newblock
    \APACrefYearMonthDay{2008}{}{}.
    \newblock
    {\BBOQ}\APACrefatitle {High-energy electron beams launched into space by
      thunderstorms} {High-energy electron beams launched into space by
      thunderstorms}.{\BBCQ}
    \newblock
    \APACjournalVolNumPages{Geophysical Research Letters}{35}{2}{}.
    \PrintBackRefs{\CurrentBib}

    \bibitem [\protect \citeauthoryear {%
    {Dwyer}%
    \ \protect \BOthers {.}}{%
    {Dwyer}%
    \ \protect \BOthers {.}}{%
    {\protect \APACyear {2012}}%
    }]{%
    TGF_GND_Dwyer2012}
    \APACinsertmetastar {%
    TGF_GND_Dwyer2012}%
    \begin{APACrefauthors}%
    {Dwyer}, J\BPBI R.%
    , Schaal, M\BPBI M.%
    , Cramer, E.%
    , Arabshahi, S.%
    , Liu, N.%
    , Rassoul, H.%
    \BDBL {}Uman, M\BPBI A.%
    \end{APACrefauthors}%
    \unskip\
    \newblock
    \APACrefYearMonthDay{2012}{}{}.
    \newblock
    {\BBOQ}\APACrefatitle {Observation of a gamma-ray flash at ground level in
      association with a cloud-to-ground lightning return stroke} {Observation of a
      gamma-ray flash at ground level in association with a cloud-to-ground
      lightning return stroke}.{\BBCQ}
    \newblock
    \APACjournalVolNumPages{Journal of Geophysical Research: Space
      Physics}{117}{A10}{}.
    \PrintBackRefs{\CurrentBib}

    \bibitem [\protect \citeauthoryear {%
    {Dwyer}%
    \ \BBA {} Smith%
    }{%
    {Dwyer}%
    \ \BBA {} Smith%
    }{%
    {\protect \APACyear {2005}}%
    }]{%
    TGF_RSC_Dwyer2005}
    \APACinsertmetastar {%
    TGF_RSC_Dwyer2005}%
    \begin{APACrefauthors}%
    {Dwyer}, J\BPBI R.%
    \BCBT {}\ \BBA {} Smith, D\BPBI M.%
    \end{APACrefauthors}%
    \unskip\
    \newblock
    \APACrefYearMonthDay{2005}{}{}.
    \newblock
    {\BBOQ}\APACrefatitle {A comparison between Monte Carlo simulations of runaway
      breakdown and terrestrial gamma-ray flash observations} {A comparison between
      monte carlo simulations of runaway breakdown and terrestrial gamma-ray flash
      observations}.{\BBCQ}
    \newblock
    \APACjournalVolNumPages{Geophysical Research Letters}{32}{22}{}.
    \PrintBackRefs{\CurrentBib}

    \bibitem [\protect \citeauthoryear {%
    Fishman%
    \ \protect \BOthers {.}}{%
    Fishman%
    \ \protect \BOthers {.}}{%
    {\protect \APACyear {1994}}%
    }]{%
    TGF_BAT_Fishman1994}
    \APACinsertmetastar {%
    TGF_BAT_Fishman1994}%
    \begin{APACrefauthors}%
    Fishman, G\BPBI J.%
    , Bhat, P\BPBI N.%
    , Mallozzi, R.%
    , Horack, J\BPBI M.%
    , Koshut, T.%
    , Kouveliotou, C.%
    \BDBL {}Christian, H\BPBI J.%
    \end{APACrefauthors}%
    \unskip\
    \newblock
    \APACrefYearMonthDay{1994}{}{}.
    \newblock
    {\BBOQ}\APACrefatitle {Discovery of Intense Gamma-Ray Flashes of Atmospheric
      Origin} {Discovery of intense gamma-ray flashes of atmospheric
      origin}.{\BBCQ}
    \newblock
    \APACjournalVolNumPages{Science}{264}{5163}{1313--1316}.
    \PrintBackRefs{\CurrentBib}

    \bibitem [\protect \citeauthoryear {%
    Grefenstette%
    , Smith%
    , Hazelton%
    \BCBL {}\ \BBA {} Lopez%
    }{%
    Grefenstette%
    \ \protect \BOthers {.}}{%
    {\protect \APACyear {2009}}%
    }]{%
    TGF_RHE_Grefenstette2009}
    \APACinsertmetastar {%
    TGF_RHE_Grefenstette2009}%
    \begin{APACrefauthors}%
    Grefenstette, B\BPBI W.%
    , Smith, D\BPBI M.%
    , Hazelton, B.%
    \BCBL {}\ \BBA {} Lopez, L.%
    \end{APACrefauthors}%
    \unskip\
    \newblock
    \APACrefYearMonthDay{2009}{}{}.
    \newblock
    {\BBOQ}\APACrefatitle {First RHESSI terrestrial gamma ray flash catalog} {First
      rhessi terrestrial gamma ray flash catalog}.{\BBCQ}
    \newblock
    \APACjournalVolNumPages{Journal of Geophysical Research: Space
      Physics}{114}{A2}{}.
    \PrintBackRefs{\CurrentBib}

    \bibitem [\protect \citeauthoryear {%
    Guo%
    \ \protect \BOthers {.}}{%
    Guo%
    \ \protect \BOthers {.}}{%
    {\protect \APACyear {2020}}%
    }]{%
    GEC_SIM_Guo2020}
    \APACinsertmetastar {%
    GEC_SIM_Guo2020}%
    \begin{APACrefauthors}%
    Guo, D.%
    , Peng, W.%
    , Zhu, Y.%
    , Li, G.%
    , Liao, J.%
    , Xiong, S.%
    \BDBL {}others%
    \end{APACrefauthors}%
    \unskip\
    \newblock
    \APACrefYearMonthDay{2020}{}{}.
    \newblock
    {\BBOQ}\APACrefatitle {Energy response and in-flight background simulationfor
      GECAM} {Energy response and in-flight background simulationfor gecam}.{\BBCQ}
    \newblock
    \APACjournalVolNumPages{SCIENTIA SINICA Physica, Mechanica \&
      Astronomica}{50}{12}{129509}.
    \PrintBackRefs{\CurrentBib}

    \bibitem [\protect \citeauthoryear {%
    Gurevich%
    , Milikh%
    \BCBL {}\ \BBA {} Roussel-Dupre%
    }{%
    Gurevich%
    \ \protect \BOthers {.}}{%
    {\protect \APACyear {1992}}%
    }]{%
    TGF_TEO_Gurevich1992}
    \APACinsertmetastar {%
    TGF_TEO_Gurevich1992}%
    \begin{APACrefauthors}%
    Gurevich, A.%
    , Milikh, G.%
    \BCBL {}\ \BBA {} Roussel-Dupre, R.%
    \end{APACrefauthors}%
    \unskip\
    \newblock
    \APACrefYearMonthDay{1992}{}{}.
    \newblock
    {\BBOQ}\APACrefatitle {Runaway electron mechanism of air breakdown and
      preconditioning during a thunderstorm} {Runaway electron mechanism of air
      breakdown and preconditioning during a thunderstorm}.{\BBCQ}
    \newblock
    \APACjournalVolNumPages{Physics Letters A}{165}{5-6}{463--468}.
    \PrintBackRefs{\CurrentBib}

    \bibitem [\protect \citeauthoryear {%
    Han%
    \ \protect \BOthers {.}}{%
    Han%
    \ \protect \BOthers {.}}{%
    {\protect \APACyear {2020}}%
    }]{%
    GEC_INS_Han2020}
    \APACinsertmetastar {%
    GEC_INS_Han2020}%
    \begin{APACrefauthors}%
    Han, X.%
    , Zhang, K.%
    , Huang, J.%
    , YU, J.%
    , XIONG, S.%
    , CHEN, Y.%
    \BDBL {}GENG, H.%
    \end{APACrefauthors}%
    \unskip\
    \newblock
    \APACrefYearMonthDay{2020}{}{}.
    \newblock
    {\BBOQ}\APACrefatitle {GECAM satellite system design and technological
      characteristic} {Gecam satellite system design and technological
      characteristic}{\BBCQ}\ [Journal Article].
    \newblock
    \APACjournalVolNumPages{SCIENTIA SINICA Physica, Mechanica \&
      Astronomica}{50}{1674-7275}{129507}.
    \newblock
    \begin{APACrefURL} \url{https://www.sciengine.com/publisher/Science China
      Press/journal/SCIENTIA SINICA Physica, Mechanica \&
      Astronomica/50/12/10.1360/SSPMA-2020-0120} \end{APACrefURL}
    \newblock
    \begin{APACrefDOI} \doi{https://doi.org/10.1360/SSPMA-2020-0120}
      \end{APACrefDOI}
    \PrintBackRefs{\CurrentBib}

    \bibitem [\protect \citeauthoryear {%
    {Kochkin}%
    \ \protect \BOthers {.}}{%
    {Kochkin}%
    \ \protect \BOthers {.}}{%
    {\protect \APACyear {2019}}%
    }]{%
    TGF_ASM_Kochkin2019}
    \APACinsertmetastar {%
    TGF_ASM_Kochkin2019}%
    \begin{APACrefauthors}%
    {Kochkin}, P.%
    , {{\O}stgaard}, N.%
    , {Neubert}, T.%
    , {Victor}, R.%
    , {Ullaland}, K.%
    , {Yang}, S.%
    \BDBL {}{Eyles}, C\BPBI J.%
    \end{APACrefauthors}%
    \unskip\
    \newblock
    \APACrefYearMonthDay{2019}{}{}.
    \newblock
    {\BBOQ}\APACrefatitle {{On multi-pulse TGFs observed by ASIM payload from the
      International Space Station.}} {{On multi-pulse TGFs observed by ASIM payload
      from the International Space Station.}}{\BBCQ}
    \newblock
    \BIn{} \APACrefbtitle {AGU Fall Meeting Abstracts} {Agu fall meeting
      abstracts}\ (\BVOL\ 2019, \BPG~AE43A-08).
    \PrintBackRefs{\CurrentBib}

    \bibitem [\protect \citeauthoryear {%
    {Kumar}%
    \ \BBA {} {Zhang}%
    }{%
    {Kumar}%
    \ \BBA {} {Zhang}%
    }{%
    {\protect \APACyear {2015}}%
    }]{%
    GRB_REV_Zhang2015}
    \APACinsertmetastar {%
    GRB_REV_Zhang2015}%
    \begin{APACrefauthors}%
    {Kumar}, P.%
    \BCBT {}\ \BBA {} {Zhang}, B.%
    \end{APACrefauthors}%
    \unskip\
    \newblock
    \APACrefYearMonthDay{2015}{}{}.
    \newblock
    {\BBOQ}\APACrefatitle {{The physics of gamma-ray bursts \& relativistic jets}}
      {{The physics of gamma-ray bursts \& relativistic jets}}.{\BBCQ}
    \newblock
    \APACjournalVolNumPages{Physics Reports}{561}{}{1-109}.
    \newblock
    \begin{APACrefDOI} \doi{10.1016/j.physrep.2014.09.008} \end{APACrefDOI}
    \PrintBackRefs{\CurrentBib}

    \bibitem [\protect \citeauthoryear {%
    {Li}%
    \ \protect \BOthers {.}}{%
    {Li}%
    \ \protect \BOthers {.}}{%
    {\protect \APACyear {2022}}%
    }]{%
    GEC_INS_Li2022}
    \APACinsertmetastar {%
    GEC_INS_Li2022}%
    \begin{APACrefauthors}%
    {Li}, X\BPBI Q.%
    , Wen, X\BPBI Y.%
    , An, Z\BPBI H.%
    , C., C.%
    , Chang, Z.%
    \BCBL {}\ \BBA {} Chen, G.%
    \end{APACrefauthors}%
    \unskip\
    \newblock
    \APACrefYearMonthDay{2022}{}{}.
    \newblock
    {\BBOQ}\APACrefatitle {The technology for detection of gamma-ray burst with
      GECAM satellite} {The technology for detection of gamma-ray burst with gecam
      satellite}.{\BBCQ}
    \newblock
    \APACjournalVolNumPages{Radiation Detection Technology and Methods}{}{}{}.
    \newblock
    \begin{APACrefDOI} \doi{10.1007/s41605-021-00288-z} \end{APACrefDOI}
    \PrintBackRefs{\CurrentBib}

    \bibitem [\protect \citeauthoryear {%
    Lindanger%
    \ \protect \BOthers {.}}{%
    Lindanger%
    \ \protect \BOthers {.}}{%
    {\protect \APACyear {2020}}%
    }]{%
    TGF_AGL_Lindanger2020}
    \APACinsertmetastar {%
    TGF_AGL_Lindanger2020}%
    \begin{APACrefauthors}%
    Lindanger, A.%
    , Marisaldi, M.%
    , Maiorana, C.%
    , Sarria, D.%
    , Albrechtsen, K.%
    , {\O}stgaard, N.%
    \BDBL {}others%
    \end{APACrefauthors}%
    \unskip\
    \newblock
    \APACrefYearMonthDay{2020}{}{}.
    \newblock
    {\BBOQ}\APACrefatitle {The 3rd AGILE terrestrial gamma ray flash catalog. Part
      I: Association to lightning sferics} {The 3rd agile terrestrial gamma ray
      flash catalog. part i: Association to lightning sferics}.{\BBCQ}
    \newblock
    \APACjournalVolNumPages{Journal of Geophysical Research:
      Atmospheres}{125}{11}{e31985}.
    \PrintBackRefs{\CurrentBib}

    \bibitem [\protect \citeauthoryear {%
    Liu%
    \ \BBA {} Dwyer%
    }{%
    Liu%
    \ \BBA {} Dwyer%
    }{%
    {\protect \APACyear {2013}}%
    }]{%
    TGF_TEO_Liu2013}
    \APACinsertmetastar {%
    TGF_TEO_Liu2013}%
    \begin{APACrefauthors}%
    Liu, N.%
    \BCBT {}\ \BBA {} Dwyer, J\BPBI R.%
    \end{APACrefauthors}%
    \unskip\
    \newblock
    \APACrefYearMonthDay{2013}{}{}.
    \newblock
    {\BBOQ}\APACrefatitle {Modeling terrestrial gamma ray flashes produced by
      relativistic feedback discharges} {Modeling terrestrial gamma ray flashes
      produced by relativistic feedback discharges}.{\BBCQ}
    \newblock
    \APACjournalVolNumPages{Journal of Geophysical Research: Space
      Physics}{118}{5}{2359--2376}.
    \PrintBackRefs{\CurrentBib}

    \bibitem [\protect \citeauthoryear {%
    {Liu}%
    \ \protect \BOthers {.}}{%
    {Liu}%
    \ \protect \BOthers {.}}{%
    {\protect \APACyear {2022}}%
    }]{%
    TGF_TEO_Liu2022}
    \APACinsertmetastar {%
    TGF_TEO_Liu2022}%
    \begin{APACrefauthors}%
    {Liu}, N\BPBI Y.%
    , {Scholten}, O.%
    , {Hare}, B\BPBI M.%
    , {Dwyer}, J\BPBI R.%
    , {Sterpka}, C\BPBI F.%
    , {Kolma{\v{s}}ov{\'a}}, I.%
    \BCBL {}\ \BBA {} {Santol{\'\i}k}, O.%
    \end{APACrefauthors}%
    \unskip\
    \newblock
    \APACrefYearMonthDay{2022}{{\APACmonth{03}}}{}.
    \newblock
    {\BBOQ}\APACrefatitle {{LOFAR Observations of Lightning Initial Breakdown
      Pulses}} {{LOFAR Observations of Lightning Initial Breakdown Pulses}}.{\BBCQ}
    \newblock
    \APACjournalVolNumPages{Geophysical Research Letters}{49}{6}{e98073}.
    \newblock
    \begin{APACrefDOI} \doi{10.1029/2022GL098073} \end{APACrefDOI}
    \PrintBackRefs{\CurrentBib}

    \bibitem [\protect \citeauthoryear {%
    Liu%
    \ \protect \BOthers {.}}{%
    Liu%
    \ \protect \BOthers {.}}{%
    {\protect \APACyear {2021}}%
    }]{%
    GEC_INS_Liu2021}
    \APACinsertmetastar {%
    GEC_INS_Liu2021}%
    \begin{APACrefauthors}%
    Liu, Y\BPBI Q.%
    , Gong, K.%
    , Li, X\BPBI Q.%
    , Wen, X\BPBI Y.%
    , An, Z\BPBI H.%
    , Cai, C.%
    \BDBL {}others%
    \end{APACrefauthors}%
    \unskip\
    \newblock
    \APACrefYearMonthDay{2021}{}{}.
    \newblock
    {\BBOQ}\APACrefatitle {The SiPM Array Data Acquisition Algorithm Applied to the
      GECAM Satellite Payload} {The sipm array data acquisition algorithm applied
      to the gecam satellite payload}.{\BBCQ}
    \newblock
    \APACjournalVolNumPages{arXiv preprint arXiv:2112.04786}{}{}{}.
    \PrintBackRefs{\CurrentBib}

    \bibitem [\protect \citeauthoryear {%
    Lu%
    \ \protect \BOthers {.}}{%
    Lu%
    \ \protect \BOthers {.}}{%
    {\protect \APACyear {2010}}%
    }]{%
    TGF_VLF_Lu2010}
    \APACinsertmetastar {%
    TGF_VLF_Lu2010}%
    \begin{APACrefauthors}%
    Lu, G.%
    , Blakeslee, R\BPBI J.%
    , Li, J.%
    , Smith, D\BPBI M.%
    , Shao, X\BHBI M.%
    , McCaul, E\BPBI W.%
    \BDBL {}Cummer, S\BPBI A.%
    \end{APACrefauthors}%
    \unskip\
    \newblock
    \APACrefYearMonthDay{2010}{}{}.
    \newblock
    {\BBOQ}\APACrefatitle {Lightning mapping observation of a terrestrial gamma-ray
      flash} {Lightning mapping observation of a terrestrial gamma-ray
      flash}.{\BBCQ}
    \newblock
    \APACjournalVolNumPages{Geophysical Research Letters}{37}{11}{}.
    \PrintBackRefs{\CurrentBib}

    \bibitem [\protect \citeauthoryear {%
    Lu%
    \ \protect \BOthers {.}}{%
    Lu%
    \ \protect \BOthers {.}}{%
    {\protect \APACyear {2011}}%
    }]{%
    TGF_VLF_Lu2011}
    \APACinsertmetastar {%
    TGF_VLF_Lu2011}%
    \begin{APACrefauthors}%
    Lu, G.%
    , Cummer, S\BPBI A.%
    , Li, J.%
    , Han, F.%
    , Smith, D\BPBI M.%
    \BCBL {}\ \BBA {} Grefenstette, B\BPBI W.%
    \end{APACrefauthors}%
    \unskip\
    \newblock
    \APACrefYearMonthDay{2011}{}{}.
    \newblock
    {\BBOQ}\APACrefatitle {Characteristics of broadband lightning emissions
      associated with terrestrial gamma ray flashes} {Characteristics of broadband
      lightning emissions associated with terrestrial gamma ray flashes}.{\BBCQ}
    \newblock
    \APACjournalVolNumPages{Journal of Geophysical Research: Space
      Physics}{116}{A3}{}.
    \PrintBackRefs{\CurrentBib}

    \bibitem [\protect \citeauthoryear {%
    Lyu%
    , Cummer%
    , Lu%
    , Zhou%
    \BCBL {}\ \BBA {} Weinert%
    }{%
    Lyu%
    \ \protect \BOthers {.}}{%
    {\protect \APACyear {2016}}%
    }]{%
    TGF_LED_Lyu2016}
    \APACinsertmetastar {%
    TGF_LED_Lyu2016}%
    \begin{APACrefauthors}%
    Lyu, F.%
    , Cummer, S\BPBI A.%
    , Lu, G.%
    , Zhou, X.%
    \BCBL {}\ \BBA {} Weinert, J.%
    \end{APACrefauthors}%
    \unskip\
    \newblock
    \APACrefYearMonthDay{2016}{}{}.
    \newblock
    {\BBOQ}\APACrefatitle {Imaging lightning intracloud initial stepped leaders by
      low-frequency interferometric lightning mapping array} {Imaging lightning
      intracloud initial stepped leaders by low-frequency interferometric lightning
      mapping array}.{\BBCQ}
    \newblock
    \APACjournalVolNumPages{Geophysical Research Letters}{43}{10}{5516--5523}.
    \PrintBackRefs{\CurrentBib}

    \bibitem [\protect \citeauthoryear {%
    {Mailyan}%
    \ \protect \BOthers {.}}{%
    {Mailyan}%
    \ \protect \BOthers {.}}{%
    {\protect \APACyear {2020}}%
    }]{%
    TGF_GBM_Mailyan2020}
    \APACinsertmetastar {%
    TGF_GBM_Mailyan2020}%
    \begin{APACrefauthors}%
    {Mailyan}, B\BPBI G.%
    , {Nag}, A.%
    , {Dwyer}, J\BPBI R.%
    , {Said}, R\BPBI K.%
    , {Briggs}, M\BPBI S.%
    , {Roberts}, O\BPBI J.%
    \BDBL {}{Rassoul}, H\BPBI K.%
    \end{APACrefauthors}%
    \unskip\
    \newblock
    \APACrefYearMonthDay{2020}{{\APACmonth{04}}}{}.
    \newblock
    {\BBOQ}\APACrefatitle {{Gamma-Ray and Radio-Frequency Radiation from
      Thunderstorms Observed from Space and Ground}} {{Gamma-Ray and
      Radio-Frequency Radiation from Thunderstorms Observed from Space and
      Ground}}.{\BBCQ}
    \newblock
    \APACjournalVolNumPages{Scientific Reports}{10}{}{7286}.
    \newblock
    \begin{APACrefDOI} \doi{10.1038/s41598-020-63437-2} \end{APACrefDOI}
    \PrintBackRefs{\CurrentBib}

    \bibitem [\protect \citeauthoryear {%
    Maiorana%
    \ \protect \BOthers {.}}{%
    Maiorana%
    \ \protect \BOthers {.}}{%
    {\protect \APACyear {2020}}%
    }]{%
    TGF_AGL_Maiorana2020}
    \APACinsertmetastar {%
    TGF_AGL_Maiorana2020}%
    \begin{APACrefauthors}%
    Maiorana, C.%
    , Marisaldi, M.%
    , Lindanger, A.%
    , {\O}stgaard, N.%
    , Ursi, A.%
    , Sarria, D.%
    \BDBL {}others%
    \end{APACrefauthors}%
    \unskip\
    \newblock
    \APACrefYearMonthDay{2020}{}{}.
    \newblock
    {\BBOQ}\APACrefatitle {The 3rd AGILE terrestrial gamma-ray flashes catalog.
      Part II: Optimized selection criteria and characteristics of the new sample}
      {The 3rd agile terrestrial gamma-ray flashes catalog. part ii: Optimized
      selection criteria and characteristics of the new sample}.{\BBCQ}
    \newblock
    \APACjournalVolNumPages{Journal of Geophysical Research:
      Atmospheres}{125}{11}{e31986}.
    \PrintBackRefs{\CurrentBib}

    \bibitem [\protect \citeauthoryear {%
    Marisaldi%
    \ \protect \BOthers {.}}{%
    Marisaldi%
    \ \protect \BOthers {.}}{%
    {\protect \APACyear {2010}}%
    }]{%
    TGF_AGL_Marisaldi2010}
    \APACinsertmetastar {%
    TGF_AGL_Marisaldi2010}%
    \begin{APACrefauthors}%
    Marisaldi, M.%
    , Fuschino, F.%
    , Labanti, C.%
    , Galli, M.%
    , Longo, F.%
    , Del~Monte, E.%
    \BDBL {}others%
    \end{APACrefauthors}%
    \unskip\
    \newblock
    \APACrefYearMonthDay{2010}{}{}.
    \newblock
    {\BBOQ}\APACrefatitle {Detection of terrestrial gamma ray flashes up to 40 MeV
      by the AGILE satellite} {Detection of terrestrial gamma ray flashes up to 40
      mev by the agile satellite}.{\BBCQ}
    \newblock
    \APACjournalVolNumPages{Journal of Geophysical Research: Space
      Physics}{115}{A3}{}.
    \PrintBackRefs{\CurrentBib}

    \bibitem [\protect \citeauthoryear {%
    Marisaldi%
    \ \protect \BOthers {.}}{%
    Marisaldi%
    \ \protect \BOthers {.}}{%
    {\protect \APACyear {2019}}%
    }]{%
    TGF_AGL_Marisaldi2019}
    \APACinsertmetastar {%
    TGF_AGL_Marisaldi2019}%
    \begin{APACrefauthors}%
    Marisaldi, M.%
    , Galli, M.%
    , Labanti, C.%
    , {\O}stgaard, N.%
    , Sarria, D.%
    , Cummer, S.%
    \BDBL {}others%
    \end{APACrefauthors}%
    \unskip\
    \newblock
    \APACrefYearMonthDay{2019}{}{}.
    \newblock
    {\BBOQ}\APACrefatitle {On the high-energy spectral component and fine time
      structure of terrestrial gamma ray flashes} {On the high-energy spectral
      component and fine time structure of terrestrial gamma ray flashes}.{\BBCQ}
    \newblock
    \APACjournalVolNumPages{Journal of Geophysical Research:
      Atmospheres}{124}{14}{7484--7497}.
    \PrintBackRefs{\CurrentBib}

    \bibitem [\protect \citeauthoryear {%
    Moss%
    , Pasko%
    , Liu%
    \BCBL {}\ \BBA {} Veronis%
    }{%
    Moss%
    \ \protect \BOthers {.}}{%
    {\protect \APACyear {2006}}%
    }]{%
    TGF_TEO_Moss2006}
    \APACinsertmetastar {%
    TGF_TEO_Moss2006}%
    \begin{APACrefauthors}%
    Moss, G\BPBI D.%
    , Pasko, V\BPBI P.%
    , Liu, N.%
    \BCBL {}\ \BBA {} Veronis, G.%
    \end{APACrefauthors}%
    \unskip\
    \newblock
    \APACrefYearMonthDay{2006}{}{}.
    \newblock
    {\BBOQ}\APACrefatitle {Monte Carlo model for analysis of thermal runaway
      electrons in streamer tips in transient luminous events and streamer zones of
      lightning leaders} {Monte carlo model for analysis of thermal runaway
      electrons in streamer tips in transient luminous events and streamer zones of
      lightning leaders}.{\BBCQ}
    \newblock
    \APACjournalVolNumPages{Journal of Geophysical Research: Space
      Physics}{111}{A2}{}.
    \PrintBackRefs{\CurrentBib}

    \bibitem [\protect \citeauthoryear {%
    {\O}stgaard%
    \ \protect \BOthers {.}}{%
    {\O}stgaard%
    \ \protect \BOthers {.}}{%
    {\protect \APACyear {2019}}%
    }]{%
    TGF_ASM_Ostgaard2019}
    \APACinsertmetastar {%
    TGF_ASM_Ostgaard2019}%
    \begin{APACrefauthors}%
    {\O}stgaard, N.%
    , Neubert, T.%
    , Reglero, V.%
    , Ullaland, K.%
    , Yang, S.%
    , Genov, G.%
    \BDBL {}others%
    \end{APACrefauthors}%
    \unskip\
    \newblock
    \APACrefYearMonthDay{2019}{}{}.
    \newblock
    {\BBOQ}\APACrefatitle {First 10 months of TGF observations by ASIM} {First 10
      months of tgf observations by asim}.{\BBCQ}
    \newblock
    \APACjournalVolNumPages{Journal of Geophysical Research:
      Atmospheres}{124}{24}{14024--14036}.
    \PrintBackRefs{\CurrentBib}

    \bibitem [\protect \citeauthoryear {%
    Poelman%
    , Schulz%
    \BCBL {}\ \BBA {} Vergeiner%
    }{%
    Poelman%
    \ \protect \BOthers {.}}{%
    {\protect \APACyear {2013}}%
    }]{%
    TGF_GLD_Poelman2013}
    \APACinsertmetastar {%
    TGF_GLD_Poelman2013}%
    \begin{APACrefauthors}%
    Poelman, D\BPBI R.%
    , Schulz, W.%
    \BCBL {}\ \BBA {} Vergeiner, C.%
    \end{APACrefauthors}%
    \unskip\
    \newblock
    \APACrefYearMonthDay{2013}{}{}.
    \newblock
    {\BBOQ}\APACrefatitle {Performance characteristics of distinct lightning
      detection networks covering Belgium} {Performance characteristics of distinct
      lightning detection networks covering belgium}.{\BBCQ}
    \newblock
    \APACjournalVolNumPages{Journal of Atmospheric and Oceanic
      Technology}{30}{5}{942--951}.
    \PrintBackRefs{\CurrentBib}

    \bibitem [\protect \citeauthoryear {%
    Pohjola%
    \ \BBA {} M{\"a}kel{\"a}%
    }{%
    Pohjola%
    \ \BBA {} M{\"a}kel{\"a}%
    }{%
    {\protect \APACyear {2013}}%
    }]{%
    TGF_GLD_Pohjola2013}
    \APACinsertmetastar {%
    TGF_GLD_Pohjola2013}%
    \begin{APACrefauthors}%
    Pohjola, H.%
    \BCBT {}\ \BBA {} M{\"a}kel{\"a}, A.%
    \end{APACrefauthors}%
    \unskip\
    \newblock
    \APACrefYearMonthDay{2013}{}{}.
    \newblock
    {\BBOQ}\APACrefatitle {The comparison of GLD360 and EUCLID lightning location
      systems in Europe} {The comparison of gld360 and euclid lightning location
      systems in europe}.{\BBCQ}
    \newblock
    \APACjournalVolNumPages{Atmospheric research}{123}{}{117--128}.
    \PrintBackRefs{\CurrentBib}

    \bibitem [\protect \citeauthoryear {%
    Roberts%
    \ \protect \BOthers {.}}{%
    Roberts%
    \ \protect \BOthers {.}}{%
    {\protect \APACyear {2018}}%
    }]{%
    TGF_GBM_Roberts2018}
    \APACinsertmetastar {%
    TGF_GBM_Roberts2018}%
    \begin{APACrefauthors}%
    Roberts, O\BPBI J.%
    , Fitzpatrick, G.%
    , Stanbro, M.%
    , McBreen, S.%
    , Briggs, M\BPBI S.%
    , Holzworth, R\BPBI H.%
    \BDBL {}Mailyan, B\BPBI G.%
    \end{APACrefauthors}%
    \unskip\
    \newblock
    \APACrefYearMonthDay{2018}{}{}.
    \newblock
    {\BBOQ}\APACrefatitle {The First Fermi-GBM Terrestrial Gamma Ray Flash Catalog}
      {The first fermi-gbm terrestrial gamma ray flash catalog}.{\BBCQ}
    \newblock
    \APACjournalVolNumPages{Journal of Geophysical Research: Space
      Physics}{123}{5}{4381-4401}.
    \newblock
    \begin{APACrefURL}
      \url{https://agupubs.onlinelibrary.wiley.com/doi/abs/10.1029/2017JA024837}
      \end{APACrefURL}
    \newblock
    \begin{APACrefDOI} \doi{https://doi.org/10.1029/2017JA024837} \end{APACrefDOI}
    \PrintBackRefs{\CurrentBib}

    \bibitem [\protect \citeauthoryear {%
    Said%
    , Cohen%
    \BCBL {}\ \BBA {} Inan%
    }{%
    Said%
    \ \protect \BOthers {.}}{%
    {\protect \APACyear {2013}}%
    }]{%
    TGF_GLD_Said2013}
    \APACinsertmetastar {%
    TGF_GLD_Said2013}%
    \begin{APACrefauthors}%
    Said, R.%
    , Cohen, M.%
    \BCBL {}\ \BBA {} Inan, U.%
    \end{APACrefauthors}%
    \unskip\
    \newblock
    \APACrefYearMonthDay{2013}{}{}.
    \newblock
    {\BBOQ}\APACrefatitle {Highly intense lightning over the oceans: Estimated peak
      currents from global GLD360 observations} {Highly intense lightning over the
      oceans: Estimated peak currents from global gld360 observations}.{\BBCQ}
    \newblock
    \APACjournalVolNumPages{Journal of Geophysical Research:
      Atmospheres}{118}{13}{6905--6915}.
    \PrintBackRefs{\CurrentBib}

    \bibitem [\protect \citeauthoryear {%
    Sarria%
    \ \protect \BOthers {.}}{%
    Sarria%
    \ \protect \BOthers {.}}{%
    {\protect \APACyear {2019}}%
    }]{%
    TEB_ASM_Sarria2019}
    \APACinsertmetastar {%
    TEB_ASM_Sarria2019}%
    \begin{APACrefauthors}%
    Sarria, D.%
    , Kochkin, P.%
    , {\O}stgaard, N.%
    , Lehtinen, N.%
    , Mezentsev, A.%
    , Marisaldi, M.%
    \BDBL {}others%
    \end{APACrefauthors}%
    \unskip\
    \newblock
    \APACrefYearMonthDay{2019}{}{}.
    \newblock
    {\BBOQ}\APACrefatitle {The first terrestrial electron beam observed by the
      atmosphere-space interactions monitor} {The first terrestrial electron beam
      observed by the atmosphere-space interactions monitor}.{\BBCQ}
    \newblock
    \APACjournalVolNumPages{Journal of Geophysical Research: Space
      Physics}{124}{12}{10497--10511}.
    \PrintBackRefs{\CurrentBib}

    \bibitem [\protect \citeauthoryear {%
    Scargle%
    , Norris%
    , Jackson%
    \BCBL {}\ \BBA {} Chiang%
    }{%
    Scargle%
    \ \protect \BOthers {.}}{%
    {\protect \APACyear {2013}}%
    }]{%
    STAT_BayesianBlock}
    \APACinsertmetastar {%
    STAT_BayesianBlock}%
    \begin{APACrefauthors}%
    Scargle, J\BPBI D.%
    , Norris, J\BPBI P.%
    , Jackson, B.%
    \BCBL {}\ \BBA {} Chiang, J.%
    \end{APACrefauthors}%
    \unskip\
    \newblock
    \APACrefYearMonthDay{2013}{feb}{}.
    \newblock
    {\BBOQ}\APACrefatitle {Studies In Astronomical Time Series Analysis. Vi.
      Bayesian Block Representations} {Studies in astronomical time series
      analysis. vi. bayesian block representations}.{\BBCQ}
    \newblock
    \APACjournalVolNumPages{The Astrophysical Journal}{764}{2}{167}.
    \newblock
    \begin{APACrefURL} \url{https://doi.org/10.1088/0004-637x/764/2/167}
      \end{APACrefURL}
    \newblock
    \begin{APACrefDOI} \doi{10.1088/0004-637x/764/2/167} \end{APACrefDOI}
    \PrintBackRefs{\CurrentBib}

    \bibitem [\protect \citeauthoryear {%
    Skeltved%
    , {\O}stgaard%
    , Mezentsev%
    , Lehtinen%
    \BCBL {}\ \BBA {} Carlson%
    }{%
    Skeltved%
    \ \protect \BOthers {.}}{%
    {\protect \APACyear {2017}}%
    }]{%
    TGF_TEO_Skeltved2017}
    \APACinsertmetastar {%
    TGF_TEO_Skeltved2017}%
    \begin{APACrefauthors}%
    Skeltved, A\BPBI B.%
    , {\O}stgaard, N.%
    , Mezentsev, A.%
    , Lehtinen, N.%
    \BCBL {}\ \BBA {} Carlson, B.%
    \end{APACrefauthors}%
    \unskip\
    \newblock
    \APACrefYearMonthDay{2017}{}{}.
    \newblock
    {\BBOQ}\APACrefatitle {Constraints to do realistic modeling of the electric
      field ahead of the tip of a lightning leader} {Constraints to do realistic
      modeling of the electric field ahead of the tip of a lightning
      leader}.{\BBCQ}
    \newblock
    \APACjournalVolNumPages{Journal of Geophysical Research:
      Atmospheres}{122}{15}{8120--8134}.
    \PrintBackRefs{\CurrentBib}

    \bibitem [\protect \citeauthoryear {%
    Stolzenburg%
    , Marshall%
    , Karunarathne%
    \BCBL {}\ \BBA {} Orville%
    }{%
    Stolzenburg%
    \ \protect \BOthers {.}}{%
    {\protect \APACyear {2016}}%
    }]{%
    TGF_LIG_Stolzenburg2016}
    \APACinsertmetastar {%
    TGF_LIG_Stolzenburg2016}%
    \begin{APACrefauthors}%
    Stolzenburg, M.%
    , Marshall, T\BPBI C.%
    , Karunarathne, S.%
    \BCBL {}\ \BBA {} Orville, R\BPBI E.%
    \end{APACrefauthors}%
    \unskip\
    \newblock
    \APACrefYearMonthDay{2016}{}{}.
    \newblock
    {\BBOQ}\APACrefatitle {Luminosity with intracloud-type lightning initial
      breakdown pulses and terrestrial gamma-ray flash candidates} {Luminosity with
      intracloud-type lightning initial breakdown pulses and terrestrial gamma-ray
      flash candidates}.{\BBCQ}
    \newblock
    \APACjournalVolNumPages{Journal of Geophysical Research:
      Atmospheres}{121}{18}{10--919}.
    \PrintBackRefs{\CurrentBib}

    \bibitem [\protect \citeauthoryear {%
    Ursi%
    , Guidorzi%
    , Marisaldi%
    , Sarria%
    \BCBL {}\ \BBA {} Frontera%
    }{%
    Ursi%
    \ \protect \BOthers {.}}{%
    {\protect \APACyear {2017}}%
    }]{%
    TGF_SAX_Ursi2017}
    \APACinsertmetastar {%
    TGF_SAX_Ursi2017}%
    \begin{APACrefauthors}%
    Ursi, A.%
    , Guidorzi, C.%
    , Marisaldi, M.%
    , Sarria, D.%
    \BCBL {}\ \BBA {} Frontera, F.%
    \end{APACrefauthors}%
    \unskip\
    \newblock
    \APACrefYearMonthDay{2017}{}{}.
    \newblock
    {\BBOQ}\APACrefatitle {Terrestrial gamma-ray flashes in the BeppoSAX data
      archive} {Terrestrial gamma-ray flashes in the bepposax data archive}.{\BBCQ}
    \newblock
    \APACjournalVolNumPages{Journal of Atmospheric and Solar-Terrestrial
      Physics}{156}{}{50--56}.
    \PrintBackRefs{\CurrentBib}

    \bibitem [\protect \citeauthoryear {%
    \APACcitebtitle {Vaisala}}{%
    \APACcitebtitle {Vaisala}}{%
    {\protect \APACyear {2022}}%
    }]{%
    YZ_TGF_GEC_SupDat_02}
    \APACinsertmetastar {%
    YZ_TGF_GEC_SupDat_02}%
    \APACrefbtitle {Vaisala.} {Vaisala.}
    \newblock
    \APACrefYearMonthDay{2022}{}{}.
    \newblock
    \APACrefnote{Vaisala. (2022). Global lightning detection network GLD360 data
      [Dataset]. Vaisala. Retrieved from
      \url{https://www.vaisala.com/en/products/systems/lightning/gld360}}
    \PrintBackRefs{\CurrentBib}

    \bibitem [\protect \citeauthoryear {%
    {Wada}%
    \ \protect \BOthers {.}}{%
    {Wada}%
    \ \protect \BOthers {.}}{%
    {\protect \APACyear {2019}}%
    }]{%
    TGF_GND_Wada2019}
    \APACinsertmetastar {%
    TGF_GND_Wada2019}%
    \begin{APACrefauthors}%
    {Wada}, Y.%
    , {Enoto}, T.%
    , {Nakamura}, Y.%
    , {Furuta}, Y.%
    , {Yuasa}, T.%
    , {Nakazawa}, K.%
    \BDBL {}{Tsuchiya}, H.%
    \end{APACrefauthors}%
    \unskip\
    \newblock
    \APACrefYearMonthDay{2019}{{\APACmonth{06}}}{}.
    \newblock
    {\BBOQ}\APACrefatitle {{Gamma-ray glow preceding downward terrestrial gamma-ray
      flash}} {{Gamma-ray glow preceding downward terrestrial gamma-ray
      flash}}.{\BBCQ}
    \newblock
    \APACjournalVolNumPages{Communications Physics}{2}{1}{67}.
    \newblock
    \begin{APACrefDOI} \doi{10.1038/s42005-019-0168-y} \end{APACrefDOI}
    \PrintBackRefs{\CurrentBib}

    \bibitem [\protect \citeauthoryear {%
    Wilson%
    }{%
    Wilson%
    }{%
    {\protect \APACyear {1925}}%
    }]{%
    TGF_TEO_Wilson1925}
    \APACinsertmetastar {%
    TGF_TEO_Wilson1925}%
    \begin{APACrefauthors}%
    Wilson, C\BPBI T.%
    \end{APACrefauthors}%
    \unskip\
    \newblock
    \APACrefYear{1925}.
    \newblock
    \APACrefbtitle {The acceleration of $\beta$-particles in strong electric fields
      such as those of thunderclouds} {The acceleration of $\beta$-particles in
      strong electric fields such as those of thunderclouds}\ (\BVOL~22).
    \PrintBackRefs{\CurrentBib}

    \bibitem [\protect \citeauthoryear {%
    {Wu}%
    \ \protect \BOthers {.}}{%
    {Wu}%
    \ \protect \BOthers {.}}{%
    {\protect \APACyear {2015}}%
    }]{%
    TGF_TEO_Wu2015}
    \APACinsertmetastar {%
    TGF_TEO_Wu2015}%
    \begin{APACrefauthors}%
    {Wu}, T.%
    , {Yoshida}, S.%
    , {Akiyama}, Y.%
    , {Stock}, M.%
    , {Ushio}, T.%
    \BCBL {}\ \BBA {} {Kawasaki}, Z.%
    \end{APACrefauthors}%
    \unskip\
    \newblock
    \APACrefYearMonthDay{2015}{{\APACmonth{09}}}{}.
    \newblock
    {\BBOQ}\APACrefatitle {{Preliminary breakdown of intracloud lightning:
      Initiation altitude, propagation speed, pulse train characteristics, and step
      length estimation}} {{Preliminary breakdown of intracloud lightning:
      Initiation altitude, propagation speed, pulse train characteristics, and step
      length estimation}}.{\BBCQ}
    \newblock
    \APACjournalVolNumPages{Journal of Geophysical Research:
      Atmospheres}{120}{18}{9071-9086}.
    \newblock
    \begin{APACrefDOI} \doi{10.1002/2015JD023546} \end{APACrefDOI}
    \PrintBackRefs{\CurrentBib}

    \bibitem [\protect \citeauthoryear {%
    {Xiao}%
    \ \protect \BOthers {.}}{%
    {Xiao}%
    \ \protect \BOthers {.}}{%
    {\protect \APACyear {2022}}%
    }]{%
    GEC_CAL_Xiao2022}
    \APACinsertmetastar {%
    GEC_CAL_Xiao2022}%
    \begin{APACrefauthors}%
    {Xiao}, S.%
    , Liu, Y.%
    , Peng, W.%
    , An, Z.%
    , Xiong, S.%
    , Tuo, Y.%
    \BDBL {}others%
    \end{APACrefauthors}%
    \unskip\
    \newblock
    \APACrefYearMonthDay{2022}{}{}.
    \newblock
    {\BBOQ}\APACrefatitle {On-ground and on-orbit time calibrations of GECAM}
      {On-ground and on-orbit time calibrations of gecam}.{\BBCQ}
    \newblock
    \APACjournalVolNumPages{Monthly Notices of the Royal Astronomical
      Society}{511}{1}{964--971}.
    \PrintBackRefs{\CurrentBib}

    \bibitem [\protect \citeauthoryear {%
    Xiong%
    \ \protect \BOthers {.}}{%
    Xiong%
    \ \protect \BOthers {.}}{%
    {\protect \APACyear {2012}}%
    }]{%
    TEB_GBM_Xiong2012}
    \APACinsertmetastar {%
    TEB_GBM_Xiong2012}%
    \begin{APACrefauthors}%
    Xiong, S.%
    , Briggs, M.%
    , Connaughton, V.%
    , Fishman, G.%
    , Tierney, D.%
    , Fitzpatrick, G.%
    \BDBL {}Hutchins, M.%
    \end{APACrefauthors}%
    \unskip\
    \newblock
    \APACrefYearMonthDay{2012}{}{}.
    \newblock
    {\BBOQ}\APACrefatitle {Location prediction of electron TGFs} {Location
      prediction of electron tgfs}.{\BBCQ}
    \newblock
    \APACjournalVolNumPages{Journal of Geophysical Research: Space
      Physics}{117}{A2}{}.
    \PrintBackRefs{\CurrentBib}

    \bibitem [\protect \citeauthoryear {%
    {Xu}%
    , Celestin%
    \BCBL {}\ \BBA {} Pasko%
    }{%
    {Xu}%
    \ \protect \BOthers {.}}{%
    {\protect \APACyear {2012}}%
    }]{%
    TGF_RSC_Xu2012}
    \APACinsertmetastar {%
    TGF_RSC_Xu2012}%
    \begin{APACrefauthors}%
    {Xu}, W.%
    , Celestin, S.%
    \BCBL {}\ \BBA {} Pasko, V\BPBI P.%
    \end{APACrefauthors}%
    \unskip\
    \newblock
    \APACrefYearMonthDay{2012}{}{}.
    \newblock
    {\BBOQ}\APACrefatitle {Source altitudes of terrestrial gamma-ray flashes
      produced by lightning leaders} {Source altitudes of terrestrial gamma-ray
      flashes produced by lightning leaders}.{\BBCQ}
    \newblock
    \APACjournalVolNumPages{Geophysical Research Letters}{39}{8}{}.
    \PrintBackRefs{\CurrentBib}

    \bibitem [\protect \citeauthoryear {%
    {Xu}%
    , Celestin%
    \BCBL {}\ \BBA {} Pasko%
    }{%
    {Xu}%
    \ \protect \BOthers {.}}{%
    {\protect \APACyear {2015}}%
    }]{%
    TGF_RSC_Xu2015}
    \APACinsertmetastar {%
    TGF_RSC_Xu2015}%
    \begin{APACrefauthors}%
    {Xu}, W.%
    , Celestin, S.%
    \BCBL {}\ \BBA {} Pasko, V\BPBI P.%
    \end{APACrefauthors}%
    \unskip\
    \newblock
    \APACrefYearMonthDay{2015}{}{}.
    \newblock
    {\BBOQ}\APACrefatitle {Optical emissions associated with terrestrial gamma ray
      flashes} {Optical emissions associated with terrestrial gamma ray
      flashes}.{\BBCQ}
    \newblock
    \APACjournalVolNumPages{Journal of Geophysical Research: Space
      Physics}{120}{2}{1355--1370}.
    \PrintBackRefs{\CurrentBib}

    \bibitem [\protect \citeauthoryear {%
    {Xu}%
    , {Celestin}%
    , {Pasko}%
    \BCBL {}\ \BBA {} {Marshall}%
    }{%
    {Xu}%
    \ \protect \BOthers {.}}{%
    {\protect \APACyear {2019}}%
    }]{%
    TGF_RSC_Xu2019}
    \APACinsertmetastar {%
    TGF_RSC_Xu2019}%
    \begin{APACrefauthors}%
    {Xu}, W.%
    , {Celestin}, S.%
    , {Pasko}, V\BPBI P.%
    \BCBL {}\ \BBA {} {Marshall}, R\BPBI A.%
    \end{APACrefauthors}%
    \unskip\
    \newblock
    \APACrefYearMonthDay{2019}{{\APACmonth{08}}}{}.
    \newblock
    {\BBOQ}\APACrefatitle {{Compton Scattering Effects on the Spectral and Temporal
      Properties of Terrestrial Gamma-Ray Flashes}} {{Compton Scattering Effects on
      the Spectral and Temporal Properties of Terrestrial Gamma-Ray
      Flashes}}.{\BBCQ}
    \newblock
    \APACjournalVolNumPages{Journal of Geophysical Research: Space
      Physics}{124}{8}{7220-7230}.
    \newblock
    \begin{APACrefDOI} \doi{10.1029/2019JA026941} \end{APACrefDOI}
    \PrintBackRefs{\CurrentBib}

    \bibitem [\protect \citeauthoryear {%
    Xu%
    \ \protect \BOthers {.}}{%
    Xu%
    \ \protect \BOthers {.}}{%
    {\protect \APACyear {2022}}%
    }]{%
    GEC_INS_Xv2021}
    \APACinsertmetastar {%
    GEC_INS_Xv2021}%
    \begin{APACrefauthors}%
    Xu, Y\BPBI B.%
    , Li, X\BPBI Q.%
    , Sun, X\BPBI L.%
    , Yang, S.%
    , Wang, H.%
    , Peng, W\BPBI X.%
    \BDBL {}Zhou, X.%
    \end{APACrefauthors}%
    \unskip\
    \newblock
    \APACrefYearMonthDay{2022}{}{}.
    \newblock
    {\BBOQ}\APACrefatitle {The design and performance of charged particle detector
      onboard the GECAM mission} {The design and performance of charged particle
      detector onboard the gecam mission}{\BBCQ}\ [Journal Article].
    \newblock
    \APACjournalVolNumPages{Radiation Detection Technology and
      Methods}{6}{1}{53-62}.
    \newblock
    \begin{APACrefURL} \url{https://doi.org/10.1007/s41605-021-00298-x}
      \end{APACrefURL}
    \newblock
    \begin{APACrefDOI} \doi{10.1007/s41605-021-00298-x} \end{APACrefDOI}
    \PrintBackRefs{\CurrentBib}

    \bibitem [\protect \citeauthoryear {%
    Zhang%
    \ \protect \BOthers {.}}{%
    Zhang%
    \ \protect \BOthers {.}}{%
    {\protect \APACyear {2022}}%
    }]{%
    GEC_INS_Zhang2022a}
    \APACinsertmetastar {%
    GEC_INS_Zhang2022a}%
    \begin{APACrefauthors}%
    Zhang, D.%
    , Li, X.%
    , Wen, X.%
    , Xiong, S.%
    , An, Z.%
    , Xu, Y.%
    \BDBL {}others%
    \end{APACrefauthors}%
    \unskip\
    \newblock
    \APACrefYearMonthDay{2022}{}{}.
    \newblock
    {\BBOQ}\APACrefatitle {Dedicated SiPM array for GRD of GECAM} {Dedicated sipm
      array for grd of gecam}.{\BBCQ}
    \newblock
    \APACjournalVolNumPages{Radiation Detection Technology and
      Methods}{6}{1}{63--69}.
    \PrintBackRefs{\CurrentBib}

    \bibitem [\protect \citeauthoryear {%
    Zhao%
    \ \protect \BOthers {.}}{%
    Zhao%
    \ \protect \BOthers {.}}{%
    {\protect \APACyear {2021}}%
    }]{%
    GEC_SFW_Yun2021}
    \APACinsertmetastar {%
    GEC_SFW_Yun2021}%
    \begin{APACrefauthors}%
    Zhao, X\BPBI Y.%
    , Xiong, S\BPBI L.%
    , Wen, X\BPBI Y.%
    , Li, X\BPBI Q.%
    , Cai, C.%
    , Xiao, S.%
    \BCBL {}\ \BBA {} Luo, Q.%
    \end{APACrefauthors}%
    \unskip\
    \newblock
    \APACrefYearMonthDay{2021}{{\APACmonth{12}}}{}.
    \newblock
    {\BBOQ}\APACrefatitle {{The In-Flight Realtime Trigger and Localization
      Software of GECAM}} {{The In-Flight Realtime Trigger and Localization
      Software of GECAM}}.{\BBCQ}
    \newblock
    \APACjournalVolNumPages{arXiv e-prints}{}{}{arXiv:2112.05101}.
    \PrintBackRefs{\CurrentBib}

    \bibitem [\protect \citeauthoryear {%
    {Zhao}%
    }{%
    {Zhao}%
    }{%
    {\protect \APACyear {2023}}%
    }]{%
    YZ_TGF_GEC_SupDat_03}
    \APACinsertmetastar {%
    YZ_TGF_GEC_SupDat_03}%
    \begin{APACrefauthors}%
    {Zhao}, Y.%
    \end{APACrefauthors}%
    \unskip\
    \newblock
    \APACrefYearMonthDay{2023}{}{}.
    \newblock
    \APACrefbtitle {Supporting Information for "The First GECAM Observation Results
      on Terrestrial Gamma-ray Flashes and Terrestrial Electron Beams".}
      {Supporting information for "the first gecam observation results on
      terrestrial gamma-ray flashes and terrestrial electron beams".}
    \newblock
    \APACrefnote{[Dataset]. Retrieved from
      \url{https://doi.org/10.5281/zenodo.8028217}}
    \PrintBackRefs{\CurrentBib}

    \bibitem [\protect \citeauthoryear {%
    Zhao%
    \ \protect \BOthers {.}}{%
    Zhao%
    \ \protect \BOthers {.}}{%
    {\protect \APACyear {2023a}}%
    }]{%
    YZ_LOC_MTD}
    \APACinsertmetastar {%
    YZ_LOC_MTD}%
    \begin{APACrefauthors}%
    Zhao, Y.%
    , Xue, W\BHBI C.%
    , Xiong, S\BHBI L.%
    , Luo, Q.%
    , Zhang, Y\BHBI Q.%
    , Yu, H.%
    \BDBL {}Zhang, S\BHBI N.%
    \end{APACrefauthors}%
    \unskip\
    \newblock
    \APACrefYearMonthDay{2023a}{}{}.
    \newblock
    {\BBOQ}\APACrefatitle {{{On} the Localization Methods of High Energy Transients
      for All-Sky Gamma-Ray Monitors}} {{{On} the Localization Methods of High
      Energy Transients for All-Sky Gamma-Ray Monitors}}.{\BBCQ}
    \newblock
    \APACjournalVolNumPages{arXiv e-prints}{}{}{arXiv:2209.13088}.
    \PrintBackRefs{\CurrentBib}

    \bibitem [\protect \citeauthoryear {%
    Zhao%
    \ \protect \BOthers {.}}{%
    Zhao%
    \ \protect \BOthers {.}}{%
    {\protect \APACyear {2023b}}%
    }]{%
    YZ_LOC_GEC}
    \APACinsertmetastar {%
    YZ_LOC_GEC}%
    \begin{APACrefauthors}%
    Zhao, Y.%
    , Xue, W\BHBI C.%
    , Xiong, S\BHBI L.%
    , Wang, Y\BHBI H.%
    , Liu, J\BHBI C.%
    , Luo, Q.%
    \BDBL {}Yu, H.%
    \end{APACrefauthors}%
    \unskip\
    \newblock
    \APACrefYearMonthDay{2023b}{}{}.
    \newblock
    {\BBOQ}\APACrefatitle {GECAM Localization of High-energy Transients and the
      Systematic Error} {Gecam localization of high-energy transients and the
      systematic error}.{\BBCQ}
    \newblock
    \APACjournalVolNumPages{The Astrophysical Journal Supplement
      Series}{265}{1}{17}.
    \newblock
    \begin{APACrefURL} \url{https://dx.doi.org/10.3847/1538-4365/acafeb}
      \end{APACrefURL}
    \newblock
    \begin{APACrefDOI} \doi{10.3847/1538-4365/acafeb} \end{APACrefDOI}
    \PrintBackRefs{\CurrentBib}

\end{thebibliography}


\end{document}


\title{Supporting Information for \\ The First GECAM Observation Results on Terrestrial Gamma-ray Flashes and Terrestrial Electron Beams}

\input{Authors_AGU}

\begin{article}

%
%

\noindent\textbf{Contents of this file}
\begin{enumerate}
\item Text S1 to Sx
\item Figures S1 to Sx
\item Tables S1 to Sx
\end{enumerate}
\noindent\textbf{Additional Supporting Information (Files uploaded separately)}
\begin{enumerate}
\item Captions for Datasets S1 to Sx
\item Captions for large Tables S1 to Sx (if larger than 1 page, upload as separate excel file)
\item Captions for Movies S1 to Sx
\item Captions for Audio S1 to Sx
\end{enumerate}

\noindent\textbf{Introduction}


\noindent\textbf{Text S1.}
%


\noindent\textbf{Data Set S1.} 


\noindent\textbf{Movie S1.} 


\noindent\textbf{Audio S1.} 
\end{article}
\clearpage